%% file: main_V5.tex
\documentclass[aps,pra,10pt,twocolumn,floatfix]{revtex4-2}

\usepackage[utf8]{inputenc}
\usepackage{times}
\usepackage{graphicx}
\usepackage{subfigure}
\usepackage{dcolumn}
\usepackage{bm}
\usepackage{xcolor}
\usepackage{amsmath,amssymb}
\usepackage{verbatim}
\usepackage{mathtools}
\usepackage{epstopdf}
\usepackage{bbold}
\usepackage{verbatim}

\usepackage{hyperref}
\hypersetup{
	colorlinks=true,
	linkcolor=blue,           
	citecolor=blue,           
	filecolor=magenta,        
	urlcolor=cyan
}

\DeclarePairedDelimiterX\braket[2]{\langle}{\rangle}{#1 \delimsize\vert #2}
\DeclarePairedDelimiterX\ketbra[2]{\lvert}{\rvert}{#1 \rangle\hspace{-.25em}\langle #2}

\begin{document}
	
	\title{Fully Non-Linear Neuromorphic Computing with Linear Wave Scattering}
	
	\author{Clara C. Wanjura}
	\address{Max Planck Institute for the Science of Light, Staudtstraße 2, 91058 Erlangen, Germany}
	
	\author{Florian Marquardt}
	\address{Max Planck Institute for the Science of Light, Staudtstraße 2, 91058 Erlangen, Germany}
	
	\date{\today}
	
	\begin{abstract}
		The increasing complexity of neural networks and the energy consumption associated with training and inference create a need for alternative neuromorphic approaches, e.g. using optics. Current proposals and implementations rely on physical non-linearities or opto-electronic conversion to realise the required non-linear activation function. However, there are significant challenges with these approaches related to power levels, control, energy-efficiency, and delays.
		
		Here, we present a scheme for a neuromorphic system that relies on linear wave scattering and yet achieves non-linear processing with a high expressivity. The key idea is to inject the input via physical parameters that affect the scattering processes. Moreover, we show that gradients needed for training can be directly measured in scattering experiments.  We predict classification accuracies on par with results obtained by standard artificial neural networks. Our proposal can be readily implemented with existing state-of-the-art, scalable platforms, e.g. in optics, microwave and electrical circuits, and we propose an integrated-photonics implementation based on racetrack resonators that achieves high connectivity with a minimal number of waveguide crossings.
	\end{abstract}
	
	
	\maketitle
	
	\section{Introduction}

	The rapid growth in neural network complexity has led to an exponential increase in energy consumption and training costs.
	This has created a need for more energy and cost efficient alternatives sparking the rapidly developing field of neuromorphic computing~\cite{markovic2020physics} in which computations are performed with physical artificial neurons.
	
	Typically, neural networks connect neurons in consecutive layers through linear maps and non-linear activation functions.
	So far, the prevalent approach has been to realise the linear maps with linear physical interactions and employ physical non-linearities (or approaches like optoelectronic conversion) to realise the non-linear activation function.
	Among the many neuromorphic computing platforms~\cite{wetzstein2020inference,shastri2021photonics,grollier2020neuromorphic,schneider2022supermind},
	optical platforms~\cite{wagner1987multilayer,shastri2021photonics,wetzstein2020inference} are one of the most promising contenders for neuromorphic computing as they efficiently allow to implement the linear aspects of the neural network offering a high degree of parallelism, high computation speeds and scalability.
	Furthermore, linear computations can be performed passively
	~\cite{bogaerts2020programmable,Harris2018Linear} while propagation losses can be very small making these devices potentially very energy efficient.
	Linear optical networks (in free space or integrated photonics) for implementing the linear aspects of neural networks, i.e., matrix-vector multiplication, are already well developed and have become the basis of commercial chips~\cite{Davies2018Loihi,bandyopadhyay2022single}.
	
	\begin{figure}[htbp]
		\centering
		\includegraphics[width=.5\textwidth]{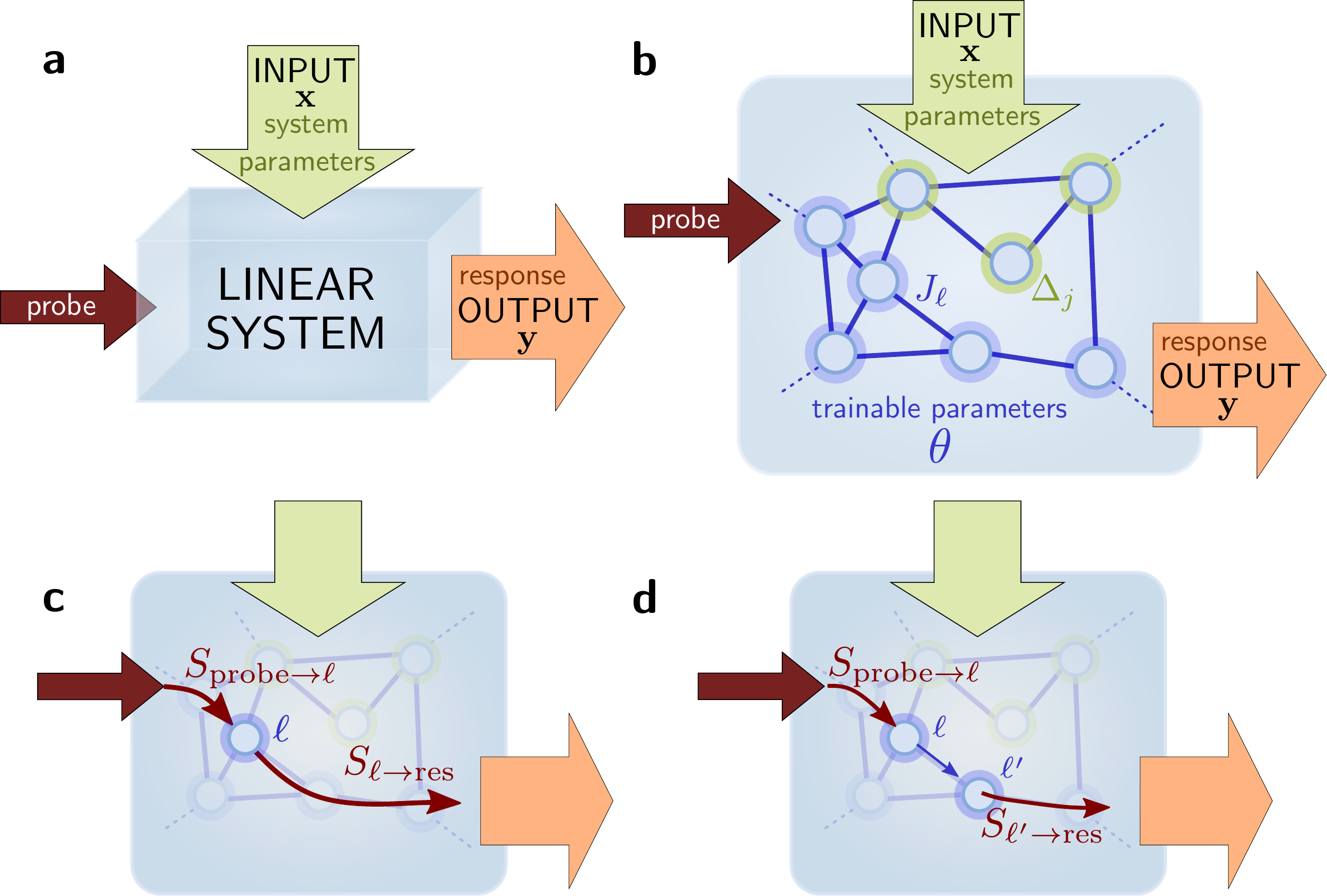}
		\caption{\textbf{Fully non-linear neuromorphic system with linear wave propagation.}
			\textbf{a}~The input to the linear, neuromorphic system is encoded in some of the tunable system parameters while the output is determined through the scattering response to a probe signal. Other controllable parameters serve as learnable parameters. 
			\textbf{b}~Example for an implementation using wave propagation through a network of coupled modes (e.g.~optical resonators).
			Here, the detunings $\Delta_j$ of some of the optical modes are utilized as input $\mathbf{x}$ while other detunings and the coupling constants $J_j$ between modes are learnable parameters $\theta$. The output is a suitable set of scattering matrix elements, Eqs.~\eqref{eq:ScatMatGeneral} and~\ref{eq:networkOutputDef}, which we obtain by comparing the response to the probe field, Eq.~\eqref{eq:ProbeResponse}.
			\textbf{c}-\textbf{d}~Due to the linearity of these systems, we have access to the gradients w.r.t.~system parameters which are required for training. The gradients are given by the products of scattering matrix elements, Eqs.~\eqref{eq:derivativeScatMatDetuning} and~\eqref{eq:derivativeScatMatCoupling}.}
		\label{fig:setup}
	\end{figure}%
	
	On the other hand, physical non-linearities to realise the non-linear aspects of the neural network are still hard to implement,
	incurring substantial hardware overhead, fabrication challenges, and other possibly demanding requirements~\cite{Zuo2019,feldmann2019all,feldmann2021parallel}, such as high laser powers in an implementation with non-linear crystals.
	Furthermore, non-linearities can induce chaotic dynamics which makes it impossible to train the system.
	An alternative approach bypasses these challenges by applying the non-linearity
	opto-electronically~\cite{hughes2018training,hamerly2019large,pai2023experimentally,chen2023deep}.
	However, proposals relying on optoelectronic conversion to realise the non-linear activation functions may be bulky, less energy-efficient and may suffer from delays.
	
	Moreover, an efficient physics-based training in the presence of non-linearities is an open challenge, although some conceptual progress has been made, especially in the form of equilibrium propagation~\cite{scellier2017equilibrium,martin2021eqspike,stern2021supervised} and Hamiltonian-echo back-propagation~\cite{lopez2021self}.
	In the presence of decay, a back-propagation algorithm was developed~\cite{guo2021backpropagation} and implemented~\cite{spall2023training} only for specific types of non-linearities.
	Another approach to physical training is to adjust parameters based on feedback~\cite{Filipovich2022Silicon} or random parameter shifts~\cite{bandyopadhyay2022single}. However, this approach
	may scale unfavourably with network size~\cite{bartunov2018assessing}.
	Other neuromorphic systems can be trained in a computer simulation~\cite{wright2022deep} which necessitate complete and accurate knowledge of the system but still offers a potential advantage in terms of speed and efficiency during inference.
	Finally, hybrid approaches~\cite{hughes2018training,pai2023experimentally} perform physical back-propagation on the linear components of the neuromorphic system.

	Here, we propose a completely new approach for a \emph{fully non-linear} neuromorphic system which is only based on \emph{linear} scattering.
	and therefore bypasses all of the challenges associated with realising and controlling physical non-linearities.
	The key idea lies in injecting the input of the neural network in the system parameters, Fig.~\ref{fig:setup}~\textbf{a}:
	While linear physical systems \emph{linearly} relate a probe signal to the response via the scattering matrix, the scattering matrix itself depends \emph{non-linearly} on system parameters.
	Therefore, we interpret some of the system parameters as the input of our neural network and some as learnable parameters.
	The output of our neural network is given by suitable scattering matrix elements which we obtain from scattering experiments by comparing the response signal to the probe signal.
	Remarkably, this non-linear dependence enables us to implement a fully functional neural network capable of performing the same tasks as standard artificial neural networks.
	As a major advantage of working with a linear scattering system,
	gradients needed to perform gradient descent can be directly measured in scattering experiments requiring only a minimal number of scattering experiments. 
	This does not require complete knowledge or control of the system; in fact, one can treat untrained parts of the system as black box.
	The neural network can be evaluated efficiently (inference) by performing a minimal number of scattering experiments promising high computation speeds.
	We simulate and train our system on hand-written digit recognition and achieve classification accuracies on par with the accuracies obtained by standard artificial neural networks.
	
	Our proposal can be readily implemented with existing state-of-the-art scalable platforms, e.g., in photonics~\cite{Harris2018Linear,bogaerts2020programmable,Pelucchi2022The} and analog electrical circuits~\cite{Indiveri2011Neuromorphic}.
	Concretely, we propose an implementation of our network with optical racetrack resonators coupled via microring resonators allowing for a densely connected network at a minimal number of waveguide crossings. 
	The architecture of our racetrack resonator array may also find application in other neuromorphic computing applications that require a high connectivity.
	
	Beyond that, our results are very general and apply to any type of linear system for which measurable quantitites can be obtained from a Green’s function. For instance, in analog electrical circuits, the scattering matrix can be replaced by the impedance matrix~\cite{Wu2004Theory,Cernanova2014Non,Cserti2011Uniform} which is obtained from the circuit's Green's function, and tunable resistances, capacitors or inductances can serve as input of the network.
	
	\section{Non-linear neuromorphic computing with linear scattering}
	
	\subsection{Concept}
	
	Waves propagating through a 
	linear multi-mode systems establish a linear relation between probe signals $\textbf{a}_\mathrm{probe}$
	and the response $\textbf{a}_\mathrm{res}$ (both are vectors in a multi-port device) via the scattering matrix
	\begin{align}\label{eq:ProbeResponse}
		\mathbf{a}_\mathrm{res}(\omega) & = S(\omega,\textbf{x},\theta) 	\mathbf{a}_\mathrm{probe}(\omega).
	\end{align}
	This matrix explicitly depends on the frequency $\omega$ of the probe signal and the system parameters, some of which we label as $\mathbf{x}$ and some as $\theta$.
	Later, $\mathbf{x}$ will represent the input vector to the neuromorphic system and $\theta$ the set of training parameters.
	
	Concretely, the system could, for instance, consist of a number of coupled modes $a_j$, such as the modes of optical resonators, which we call \emph{neuron modes}.
	The scattering matrix is determined by the linear evolution equations for the neuron modes $a_j$ collected in the vector $\mathbf{a}\equiv(a_1,\dots,a_N)^\mathrm{T}$.
	In the frequency domain,
	$\mathbf{a}(\omega)=\frac{1}{\sqrt{2\pi}} \int_{-\infty}^\infty \mathrm{d}t e^{\mathrm{i}\omega t} \mathbf{a}(t)$, the equations of motion take the form
	\begin{align}\label{eq:eomsGeneral}
		-\mathrm{i}\omega \mathbf{a}(\omega) & = -\mathrm{i} H(\textbf{x},\theta) \mathbf{a}(\omega) - \sqrt{\kappa}\, \mathbf{a}_\mathrm{probe}(\omega).
	\end{align}
	$H$ denotes the dynamic matrix which explicitly depends on the system parameters $\textbf{x}$ and $\theta$, and $\kappa\equiv\mathrm{diag}\,(\kappa_{1},\dots,\kappa_{N})$ are the external decay rates of the neuron modes to the probe waveguides.
	We obtain the scattering matrix from Eq.~\eqref{eq:eomsGeneral} through
	input-output boundary conditions, $a_{j,\mathrm{res}} = a_{j,\mathrm{probe}} + \sqrt{\kappa_{j}} a_j$~\cite{Gardiner1985Input,Clerk2010Introduction},
	\begin{align}\label{eq:ScatMatGeneral}
		S(\omega,\textbf{x},\theta)
		& = \mathbb{1} - \mathrm{i}  \sqrt{\kappa} G(\omega,\textbf{x},\theta)  \sqrt{\kappa} \notag \\
		& = \mathbb{1} - \mathrm{i}  \sqrt{\kappa} [\omega\mathbb{1}-H(\textbf{x},\theta)]^{-1}  \sqrt{\kappa}.
	\end{align}
	with the system's Green's function $G(\omega,\textbf{x},\theta) \equiv -\mathrm{i}(\omega\mathbb{1}-H(\omega,\textbf{x},\theta))^{-1}$.

	Eq.~\eqref{eq:ScatMatGeneral} reveals the general idea behind our concept.
	While the relation between probe and response is \emph{linear}, Eq.~\eqref{eq:ProbeResponse}, the scattering matrix $S(\omega,\mathbf{x},\theta)$, Eq.~\eqref{eq:ScatMatGeneral}, is a \emph{non-linear} function of the system parameters.
	Hence, interpreting some of the system parameters~$\mathbf{x}$ as inputs, and some as learnable parameters~$\theta$, see Fig.~\ref{fig:setup}~\textbf{a}, we are now able to represent learnable non-linear functions of the input with the help of the scattering matrix.
	
	The network's output is a suitable set of scattering matrix elements $S_{j,\ell}$ which are obtained as the ratio between the response and the probe signal. Since the scattering matrix is generally complex, one can consider real or imaginary part as output, or, more generally,
	\begin{align}\label{eq:networkOutputDef}
		y_r = \mathrm{Re}\, (e^{\mathrm{i}\phi}S_{r,p}),
	\end{align}
	with a suitable set of $p$ and $r$ and a convenient $\phi$.
	$p$ here refers to a probe site and $r$ to the sites at which the response is recorded. The number of response sites $r$ equals the output dimension $N_\mathrm{out}$.
	The phase $\phi$ allows to select a quadrature, e.g., $\phi=0$ corresponds to $\mathrm{Re}\,S_{r,p}$ whereas $\phi=\frac{\pi}{2}$ corresponds to $\mathrm{Im}\,S_{r,p}$ which we will use later on. In practise, the quadrature can be measured through homodyne detection.
	There are two obvious, convenient choices for $p$ and $r$: (i)~the probe site $p$ is fixed and we consider different sites $r$ at which we record the system response. This allows to record the output of the neuromorphic system (inference) with a single measurement. (ii)~We set $p=r$. This requires $N_\mathrm{out}$ measurements to infer the output, but we found that in this case, the training converges more reliably.
	
	\subsection{Input replication for improved non-linear expressivity}
	
	The non-linearity of the scattering matrix~\eqref{eq:ScatMatGeneral} as function of the input $\mathbf{x}$ stems from the calculation of the matrix inverse representing a specific type of non-linearity. For instance, considering a scalar input $x$, any scattering matrix element $S_{j,\ell}$ is of the form $S_{j,\ell} = (a x + b)/(c x + d)$ with $a$, $b$, $c$, $d$ some quantities that depend on the other system parameters. This non-linearity stems from the fact that waves can propagate back and forth between modes and the scattering matrix~\eqref{eq:ScatMatGeneral} is a result of the interference between all these waves.
	We can see this by expressing the matrix inverse in Eq.~\eqref{eq:ScatMatGeneral} for a specific matrix element via the adjugate, $(H^{-1})_{j,\ell}=(-1)^{j+\ell} \frac{\det H_{j,\ell}}{\det H}$, in which $H_{j,\ell}$ denotes the matrix in which the $j$th row and the $\ell$th line have been omitted. Expanding the determinant expressions with the Laplace expansion, it is straightforward to see that both $\mathrm{det}H_{j,\ell}$ and $\mathrm{det}H$ linearly depend on any matrix entry $H_{m,n}$; therefore, both denominator and numerator in the previous expression depend linearly on $x$.
	
	Note, however, that the matrix determinant of $\det H$ (and similarly $\det H_{j,\ell}$) contains terms such as $H_{1,1}H_{2,2}H_{3,3}\cdots$ and $H_{1,1}H_{2,3}H_{3,2}\cdots$. This allows us to overcome the seeming limitation of the scattering matrix,
	namely, by letting the input value $x$ explicitly enter in more than one system parameter.
	In this way, it is possible to show, that one can represent general non-linear functions with the scattering matrix in which the number of repetitions $R$ of the input determines how well the target function can be approximated. Letting $R\to\infty$, the approximation approaches the target function.
	We prove this in the SI for one-dimensional inputs.
	
	Most applications of machine learning, however, operate on high-dimensional input spaces. In this case, the situation becomes even more favourable:
	even at a single replication of the input, the scattering matrix allows to encode correlations between different elements of the input vector, i.e., the scattering matrix automatically includes terms such as $x_1 x_2 x_3 \cdots$. Therefore, in practise, it can be sufficient to replicate the input only once or twice. In particular, we will show later that for a digit recognition task, Fig.~\ref{fig:MNISTTraining}, it was sufficient to replicate the input once, $R=1$. 
	
	\subsection{Training}
	
	We will now show one particularly useful consequence of working with a linear scattering system: it is possible to perform gradient descent based on physically measurable gradients, which is rare in neuromorphic systems.
	Specifically, gradients w.r.t. $\theta$ are directly measurable as products of scattering matrix elements.
	
	The aim of the training is to minimise a cost function $\mathcal{C}$, such as the square distance between the target output $\mathbf{y}_\mathrm{tar}$ and the output $\mathbf{y}$ of the system, Eq.~\eqref{eq:networkOutputDef},
	$\mathcal{C}=\lvert \mathbf{y}_\mathrm{tar} - \mathbf{y}\rvert^2$.
	The derivative of the cost function w.r.t. a learnable parameter $\theta_j$ is given by
	\begin{align}\label{eq:DerivativeCostFunc}
		\frac{\partial \mathcal{C}}{\partial \theta_j} = \nabla_y \mathcal{C} \cdot \left(\frac{\partial \mathbf{y}}{\partial \theta_j}\right)
	\end{align}
	with $\mathbf{y}$ defined by Eq.~\eqref{eq:networkOutputDef}.
	so it only depends on the derivative of the scattering matrix.
	
	Given the mathematical structure of the scattering matrix~\eqref{eq:ScatMatGeneral}, we obtain a simple formula for its derivative 
	\begin{align}\label{eq:SMatDerivativeGeneral}
		\frac{\partial S_{r,p}}{\partial \theta_j} & = \mathrm{i}\kappa_\mathrm{p} \left(G \frac{\partial H}{\partial \theta_j}G\right)_{r,p}
	\end{align}
	with $G = \sqrt{\kappa}^{-1}(S - \mathbb{1})\sqrt{\kappa}^{-1}$ the Green's function~\eqref{eq:ScatMatGeneral},
	see the SI for the derivation.
	Since $\partial H/\partial \theta_j$ is local, this implies that gradients can be obtained from scattering experiments as combination of scattering matrix elements. Gradients can be physically extracted allowing for very efficient training. 

	We make this more concrete in the next section in which we consider an optical system whose learnable parameters are the mode detunings and coupling strengths between modes.
	
	\subsection{Implementation based on coupled modes}
	To make the previous considerations more concrete, we consider a system of
	coupled radiation modes, e.g., in the microwave or optical regime or in electrical circuits composed of coupled resonant modes, but stress that the concept is general and applies to a variety of systems and platforms.
	In an optical system, the neuron modes can be represented by cavity modes, Fig.~\ref{fig:setup}~\textbf{b}, at frequencies $\omega_{j}$ at detunings $\Delta_j\equiv\omega_{j} - \omega_0$ from some suitable reference frequency $\omega_0$. They all experience intrinsic decay at rate $\kappa'_j$.
	The modes can, in full generality, be coupled via any form of bi-linear coherent interactions, such as beam-splitter interactions, single-mode or two-mode squeezing, or via linear engineered dissipators. For simplicity, we will focus on beam-splitter couplings at strength $J_j$ as shown in Fig.~\ref{fig:setup}~\textbf{b}.
	Furthermore, a probe $a_{j,\mathrm{probe}}$ is injected via a waveguide coupled to the modes at rate $\kappa_{j}$ which induces additional losses at this rate.
	The detunings of some of the neuron modes serve as input $\textbf{x}$ while other tunable parameters, such as detunings and coupling strengths, are the learnable parameters $\theta$, see Fig.~\ref{fig:setup}~\textbf{b}. We obtain the network's output~\eqref{eq:networkOutputDef} by performing a suitable homodyne measurement on the response fields $a_{j,\mathrm{res}}$.
	Importantly, not all of the system parameters need to be tunable for our concept to work.
	
	Coming back to training, we can now more explicitly present the gradients of the scattering matrix elements defining the output~\eqref{eq:SMatDerivativeGeneral} for the parameters of this coupled-mode structure.
	In particular, the derivative w.r.t. detuning at the $j$th site is given by the scattering path from the probe site $r$ to site $j$, and from $j$ to the response site $r$ as illustrated in Fig.~\ref{fig:setup}~\textbf{c}
	\begin{align}\label{eq:derivativeScatMatDetuning}
		\frac{\partial S_{r,p}}{\partial \Delta_j} & = \mathrm{i}\frac{1}{\kappa_j} \, G_{j,p} G_{r,j}
	\end{align}
	with the Green's function $G_{m,n} = (S_{m,n} - \delta_{m,n})/\sqrt{\kappa_m\kappa_n}$.
	Similarly, the derivative w.r.t. the coupling between the $j$th and $\ell$th site is the sum of the possible scattering paths from the probe site to $j$ or $\ell$ and the response site, Fig.~\ref{fig:setup}~\textbf{d},
	\begin{align}\label{eq:derivativeScatMatCoupling}
		\frac{\partial S_{r,p}}{\partial J_{j,\ell}} & = \mathrm{i}\frac{1}{\sqrt{\kappa_j \kappa_\ell}} (G_{j,p} G_{r,\ell} + G_{\ell,p} G_{r,j}).
	\end{align}
	As we will show in more detail below, the physical extraction of gradients is most straightforward in integrated-photonics platforms based on coupled resonators~\cite{hughes2018training}, with probe waveguides attached to every resonator whose resonance frequency represents a tunable training parameter. Other, non-tunable linear components can be freely added to the setup and need not even be fully characterized or fabricated according to some specified design: they can be treated as a black box, while the training procedure will automatically take care of accounting for their effect on the wave propagation.
	Gradients can be efficiently measured requiring only $N_\mathrm{out}$ measurements with $N_\mathrm{out}$ the dimension of the network output.
	These measurements record the full scattering response to a probe at any of the probe sites $p$, so the network can be evaluated (inference) at the same time as computing the gradients.
	
	During training, gradients can either be applied directly or be post-processed to average over a mini-batch or to perform more advanced adaptive gradient-descent optimization (like with the well-known Adam optimizer). Post-processing can be directly implemented with analog electronics, e.g., to perform the sum operation over multiple gradients~\cite{pai2023experimentally}.
	
	Note that while we focus on photonic systems for the rest of this work, our results are very general and apply to any type of linear system for which measurable quantitites can be obtained from a Green’s function. In fact, linear equations of the form~\eqref{eq:eomsGeneral} appear in many different contexts so we expect our findings to have wide-ranging application. For instance, for electrical circuits, the scattering matrix can be replaced by the impedance matrix~\cite{Wu2004Theory,Cernanova2014Non,Cserti2011Uniform} which is obtained from the circuit's Green's function and tunable resistances can serve as input of the network, see the SI for more details.
	
	\begin{figure*}[htbp]
		\centering
		\includegraphics[width=\textwidth]{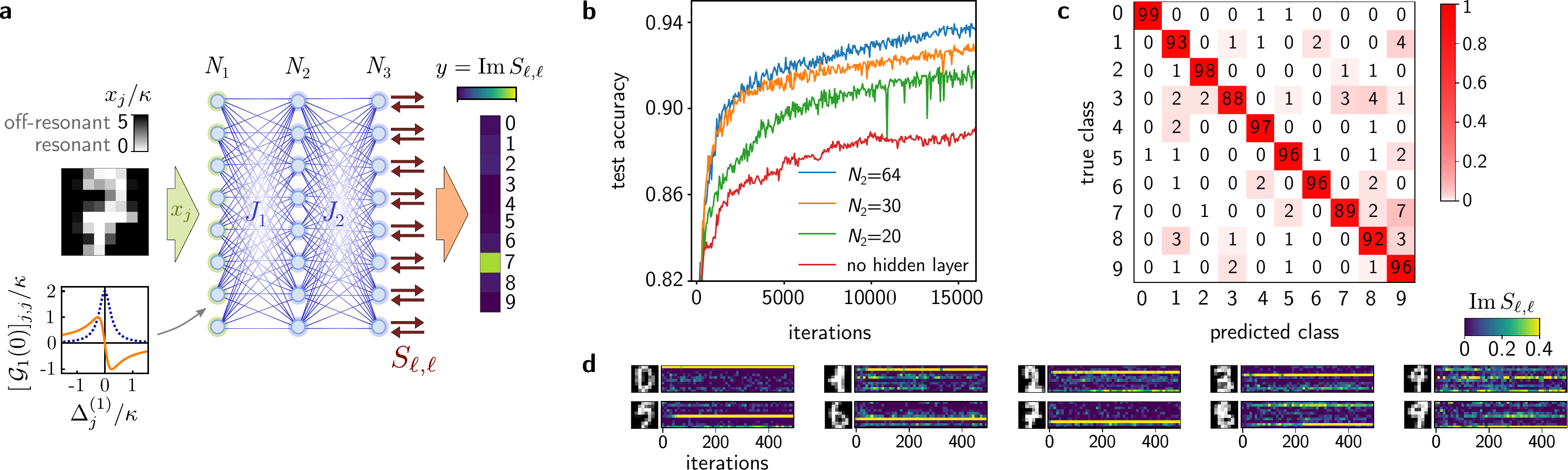}
		\caption{\textbf{Digit recognition using a scattering neuromorphic system with layered structure.}
			\textbf{a}~Scattering network used for digit recognition consisting of two or three fully connected layers with $N_1=128$, $N_3=10$ and either without hidden layer or with $N_2\in\{20,30,64\}$. We consider equal decay rates $\kappa$, set the intrinsic decay to zero $\kappa'=0$ and start from $J/\kappa=2$ with random disorder on top. The input consisting of 64 greyscale pixel values is encoded in the detuning of the first layer to which we initially add a trainable offset which is initialised according to Eq.~\eqref{eq:initialDetunings}. A vector of pixel values serves as input in which we detune the background to $x_j=5\kappa$ and make the foreground, the numerals, resonant $x_j=0$. The inset illustrates the non-linear effect of the first layer showing real and imaginary part of $[\mathcal{G}_1(0)]_{j,j}$, Eq.~\eqref{eq:recursiveGreensFunction}. The response to a probe signal at the third layer constitutes the output vector. The index of the maximal response constitutes the class.
			\textbf{b}~Evolution of the test accuracy during training for different architectures: without hidden layer, or with $N_2=\{20,30,64\}$. Here, an iteration over one mini-batch corresponds to evaluating $200$ randomly chosen images while the shown test accuracy is evaluated on the entire test set.
			Increasing the size of the hidden layer improves both the convergence speed and the best accuracy.
			\textbf{c}~The confusion matrix of the trained model with $N_2=64$ after $28,000$ iterations.
			The overall test accuracy amounts to $94.66\,\%$.
			\textbf{d}~Convergence for individual input pictures. The scattering matrix element with the largest imaginary part indicates the class. In most cases, the training rapidly converges towards the correct classification results. Digits of similar appearance, however, are frequently mistaken for the other, such as the digit 4 and 9, and only converge relatively late during training. For the training run shown here, we used Adam optimisation.
		}
		\label{fig:MNISTTraining}
	\end{figure*}%
	
	\section{Layered architecture}
	\subsection{Recursive solution to the scattering problem}
	The principle of information processing in the neuromorphic platform introduced here is fundamentally different from standard neural networks since the waves representing the information will be scattered back and forth inside the device rather than propagating unidirectionally. Nevertheless, with this in mind, we can choose an architecture that is at least inspired by the typical layer-wise structure of artificial neural networks. It allows us to gain analytic insight into the scattering matrix and to derive optimal layer sizes to make efficient use of the number of independent parameters available.
	
	We consider a layered architecture as sketched in Fig.~\ref{fig:MNISTTraining}~\textbf{a} with $L$ layers and $N_\ell$ neuron modes in the $\ell$th layer.
	Neuron modes are only coupled to neuron modes in consecutive layers but not within a layer. 
	Note that while we sketch fully connected layers in Fig.~\ref{fig:MNISTTraining}~\textbf{a}, the network does not, in principle, have to be fully connected.
	However, fully connected layers have the advantage that there is \emph{a priori} no ordering relation between modes based on their proximity within a layer.
	
	This architecture allows us to gain analytic insight into the mathematical structure of the scattering matrix.
	For a compact notation, we split the vector $\mathbf{a}$ of neuron modes, Eq.~\eqref{eq:eomsGeneral}, into vectors $\mathbf{a}_n\equiv(a_1^{(n)},\dots,a_{N_n}^{(n)})$ collecting the neuron modes of the $n$th layer.
	Correspondingly, we define the detunings of the neuron modes in the $n$th layer $\Delta^{(n)}=\mathrm{diag}\,(\Delta_1^{(n)}, \dots, \Delta_{N_n}^{(n)})$, the extrinsic decay rates to the waveguides $\kappa^{(n)}=\mathrm{diag}\,(\kappa_1^{(n)}, \dots, \kappa_{N_n}^{(n)})$, the intrinsic decay rates ${\kappa'}^{(n)}=\mathrm{diag}\,({\kappa'}_{1}^{(n)}, \dots, {\kappa'}_{N_n}^{(n)})$, the total decay rate ${\kappa_\mathrm{tot}}^{(n)}={\kappa'}^{(n)} + \kappa^{(n)}$ and $J^{(n)}$ the coupling matrix between layer $n$ and $(n+1)$ in which the element $J_{j,\ell}^{(n)}$ connects neuron mode $j$ in layer $n$ to neuron mode $\ell$ in layer $(n+1)$.
	Note that this accounts for the possibility of attaching waveguides to all modes in each layer to physically evaluate gradients according to Eqs.~\eqref{eq:derivativeScatMatDetuning} and \eqref{eq:derivativeScatMatCoupling}.
	In the frequency domain, we obtain for the equations of motion of the $n$th layer
	\begin{align}\label{eq:eomsLayers}
		-\mathrm{i}\omega \mathbf{a}_n & = \left(-\frac{{\kappa}_\mathrm{tot}^{(n)}}{2} - \mathrm{i}\Delta^{(n)}\right)\mathbf{a}_n - \mathrm{i} J^{(n)} \mathbf{a}_{n+1} - \mathrm{i} J^{(n-1)} \mathbf{a}_{n-1} \notag \\
		& \hphantom{=\left(-\frac{{\kappa'}^{(n)}}{2} - \mathrm{i}\Delta^{(n)}\right)\mathbf{a}_n}
		- \sqrt{\kappa^{(n)}} \mathbf{a}_{n,\mathrm{probe}},
	\end{align}
	in which we omitted the frequency argument $\omega$ for clarity.
	Eq.~\eqref{eq:eomsLayers} does not have the typical structure of a feed-forward neural network since neighbouring layers are coupled to the left and right---a consequence of the wave propagation through the system.
	
	We are interested in calculating the scattering response at the last layer $L$ (the output layer) which defines the output of the neuromorphic system, so we set $\mathbf{a}_{n,\mathrm{probe}}=0$ for $n\neq L$ in Eq.~\eqref{eq:eomsLayers} and only consider the response to the probe fields $\mathbf{a}_{L,\mathrm{probe}}$.
	The following procedure allows us to calculate the scattering matrix block $S_\mathrm{out}$ relating only $\mathbf{a}_{L,\mathrm{probe}}$ to $\mathbf{a}_{L,\mathrm{res}}$. A suitable set of matrix elements of $S_\mathrm{out}$ then defines the output of the neuromorphic system via Eq.~\eqref{eq:networkOutputDef}.
	
	Solving for $\mathbf{a}_1(\omega)$, then $\mathbf{a}_2(\omega)$ and subsequent layers up to $\mathbf{a}_L(\omega)$, we obtain a recursive formula for $\mathbf{a}_n(\omega)$
	\begin{align}\label{eq:neuronValues}
		\mathbf{a}_n & = \mathrm{i} \mathcal{G}_n J^{(n)} \mathbf{a}_{n+1}
	\end{align}
	with
	\begin{align}\label{eq:recursiveGreensFunction}
		\mathcal{G}_n(\omega) & \equiv \left[\frac{{\kappa}_\mathrm{tot}^{(n)}}{2}+\mathrm{i}(\Delta^{(n)}-\omega) + {J^{(n-1)}}^\dagger \mathcal{G}_{n-1} J^{(n-1)}\right]^{-1}
	\end{align}
	and $\mathcal{G}_0=0$
	so that in the last layer, we have
	\begin{align}\label{eq:ScatMatLayerNetwork}
		\mathbf{a}_L & = \mathcal{G}_L(\omega) \mathbf{a}_{L,\mathrm{probe}}.
	\end{align}
	At the last layer, the matrix $\mathcal{G}_L(\omega)$ is equal to the system's Green's function $G(\omega)=\mathcal{G}_L(\omega)$, Eq.~\eqref{eq:ScatMatGeneral}.
	
	Employing input-output boundary conditions~\cite{Clerk2010Introduction}, $\mathbf{a}_{L,\mathrm{res}} = \mathbf{a}_{L,\mathrm{probe}} + \sqrt{\kappa^{(L)}} \mathbf{a}_L$, we obtain the scattering matrix for the response at the last layer
	\begin{widetext}
		\begin{align}\label{eq:ScatMatLayers}
			S_\mathrm{out}(\omega, x, \theta) & = \mathbb{1}
			+ \sqrt{\kappa^{(L)}}
			\left[\frac{{\kappa}_\mathrm{tot}^{(L)}}{2}+\mathrm{i}(\Delta^{(L)}-\omega) + {J^{(L-1)}}^\dagger \mathcal{G}^{(L-1)} J^{(L-1)}\right]^{-1} \sqrt{\kappa^{(L)}}
		\end{align}
	\end{widetext}
	The structure of Eqs.~\eqref{eq:ScatMatLayers} and \eqref{eq:recursiveGreensFunction} is reminiscent of a generalised continued fraction, with the difference that scalar coefficients are replaced by matrices. We explore this analogy further in the SI where we also show that for scalar input and output, the scattering matrix~\eqref{eq:ScatMatLayers} can approximate arbitrary analytic functions.
	Furthermore, the recursive structure defined by Eqs.~\eqref{eq:recursiveGreensFunction} and~\eqref{eq:ScatMatLayerNetwork} mimics that of a standard artificial neural network in which the weight matrix is replaced by the coupling matrix and the matrix inverse serves as activation function. However, in contrast to the standard activation function, which is applied element-wise for each neuron, the matrix inversion acts on the entire layer.
	To gain intuition for the effect of taking the matrix inverse, we plot a diagonal entry of $[\mathcal{G}_1(0)]_{j,j}/\kappa=[\frac{1}{2}+\mathrm{i} \Delta^{(1)}_j/{\kappa}^{(1)}_j]^{-1}$ in Fig.~\ref{fig:MNISTTraining}~\textbf{a}.
	The real part follows a Lorentzian whereas the imaginary part is reminiscent of a tapered sigmoid function.
	
	The recursive structure of the scattering matrix~\eqref{eq:ScatMatLayers} bears some resemblance to the variational quantum circuits of the quantum machine learning literature~\cite{garcia2022systematic,Huang2020Quantum} in which the subsequent application of discrete unitary operators allows to realise non-linear operations. In contrast, here, we consider the steady state scattering response which allows waves to propagate back and forth giving rise to yet more complicated non-linear maps~\eqref{eq:ScatMatGeneral}.
	
	\subsection{Test case: digit recognition}
	To benchmark our model, we simulate and train a network with three layers, see Fig.~\ref{fig:MNISTTraining}~\textbf{a}, on handwritten digit classification on a down-sized version of the MNIST dataset~\cite{Alpaydin1998Optical}. The set consists of $3,823$ $8\times8$ pixels training images of numerals between $0$ and $9$ as well as $1,797$ test images.
	We choose an architecture in which the input is encoded in the detunings $\Delta^{(1)}_{j}$ of the first layer to which we add a trainable offset $\Delta^{(0)}_{j}$
	\begin{align}\label{eq:initialDetunings}
		\Delta_{2j}^{(1)} & = \Delta_{2j}^{(0)} + x_j \notag\\
		\Delta_{2j+1}^{(1)} & = - \Delta_{2j+1}^{(0)} + x_j
	\end{align}
	which we initially set to $\Delta_{2j} = \Delta_{2j+1} = 4\kappa$ (for simplicity, we consider equal decay rates $\kappa_j^{(n)}\equiv\kappa$ from now on). In this way, the input enters the second layer in the form of $[G_1(0)]_{j,j}/\kappa$ once with positive, once with negative sign, see the plot of $[G_1(0)]_{j,j}/\kappa$ in Fig.~\ref{fig:MNISTTraining}~\textbf{b}.
	We found that, beyond improving the expressivity as explained above, this also leads to faster convergence of the training.
	This doubling of the input fixes the layer size to $N_1=2\cdot 64$.
	The second layer (hidden layer) can be of variable size and we train a system (i)~without the hidden layer, (ii)~with $N_2=20$, (iii)~with $N_2=30$, (iv)~with $N_2=64$.
	We use one-hot encoding of the classes, so the output layer is fixed at $N_3=10$.
	We choose equal decay rates $\kappa$ and express all other parameters in terms of $\kappa$. For simplicity, we set the intrinsic decay to zero $\kappa'=0$.
	
	We initialise the system by making the neuron modes in the second and third layer resonant and add a small amount of disorder, i.e., $\Delta_j^{(n)}/\kappa=w_\Delta (\xi_{j}^{(n)}-1/2)$ with $w_\Delta = 0.002$ and $\xi\in[0,1)$. Similarly, we set the coupling rates to $2\kappa$ and add a small amount of disorder, $J_{j,\ell}^{(n)}/\kappa=2+w_J \xi_{j,\ell}^{(n)}$ with $w_J=0.2$. We found empirically that this initialisation leads to the fastest convergence of the training.
	Furthermore, we scale our input images such that the background is off-resonant at $\Delta/\kappa=5$ and the numerals are resonant $\Delta/\kappa=0$, see Fig.~\ref{fig:MNISTTraining}~\textbf{a} since, otherwise, the initial gradients are very small.
	
	As output of the system, we consider the imaginary part of the diagonal entries of the scattering matrix~\eqref{eq:ScatMatLayers} at the last layer at $\omega=0$, i.e. $y_\ell\equiv \mathrm{Im}\,S_{\ell,\ell}(0,x,\theta)$, Fig.~\ref{fig:MNISTTraining}~\textbf{a}.
	The goal is to minimise the cost function $\mathcal{C} = \lvert \mathbf{y}_\mathrm{tar} - \mathbf{y}\rvert^2$ in which $\mathbf{y}_\mathrm{tar}$ is the target output vector of the system which is $1$ at the index of the correct class and zero elsewhere.
	We train the system by performing gradient descent, i.e., for one mini-batch we select $200$ random images, and compute the gradients of the cost function according to which we adjust the system parameters: the detunings $\Delta_j^{(n)}$ and the coupling rates $J_{j,\ell}^{(n)}$.
	
	We show the accuracy evaluated on the test set at different stages during the training for four different architectures in Fig.~\ref{fig:MNISTTraining}~\textbf{b}. Both the convergence speed and the maximally attainable test accuracy depend on the size of the hidden layer with the system without hidden layer performing worst. This should not surprise us, since an increase in $N_2$ also increases the number of trainable parameters.
	We obtained the best classification accuracy on the test set of $94.7\%$
	for the system with $N_2=64$ after iterating over $28,000$ mini-batches.
	We show the associated confusion matrix in Fig.~\ref{fig:MNISTTraining}~\textbf{c}. This accuracy is on par with the accuracy of $94.7\%$ achieved by a standard artificial neural network.
	The artificial model we trained had an input layer of size $64$, a hidden layer of size $64$, an output layer of size $10$, and we used a sigmoid activation function.
	
	We show the evolution of the output scattering matrix elements evaluated for a few specific images in Fig.~\ref{fig:MNISTTraining}~\textbf{d}. During training, the scattering matrix converges to the correct classification result quickly. Only for such images that look very similar, e.g., the image of the $4$ and $9$ shown in Fig.~\ref{fig:MNISTTraining}~\textbf{d}, the training oscillates between the two classes and takes a longer time to converge.
	
	\begin{figure*}
		\centering
		\includegraphics[width=\textwidth]{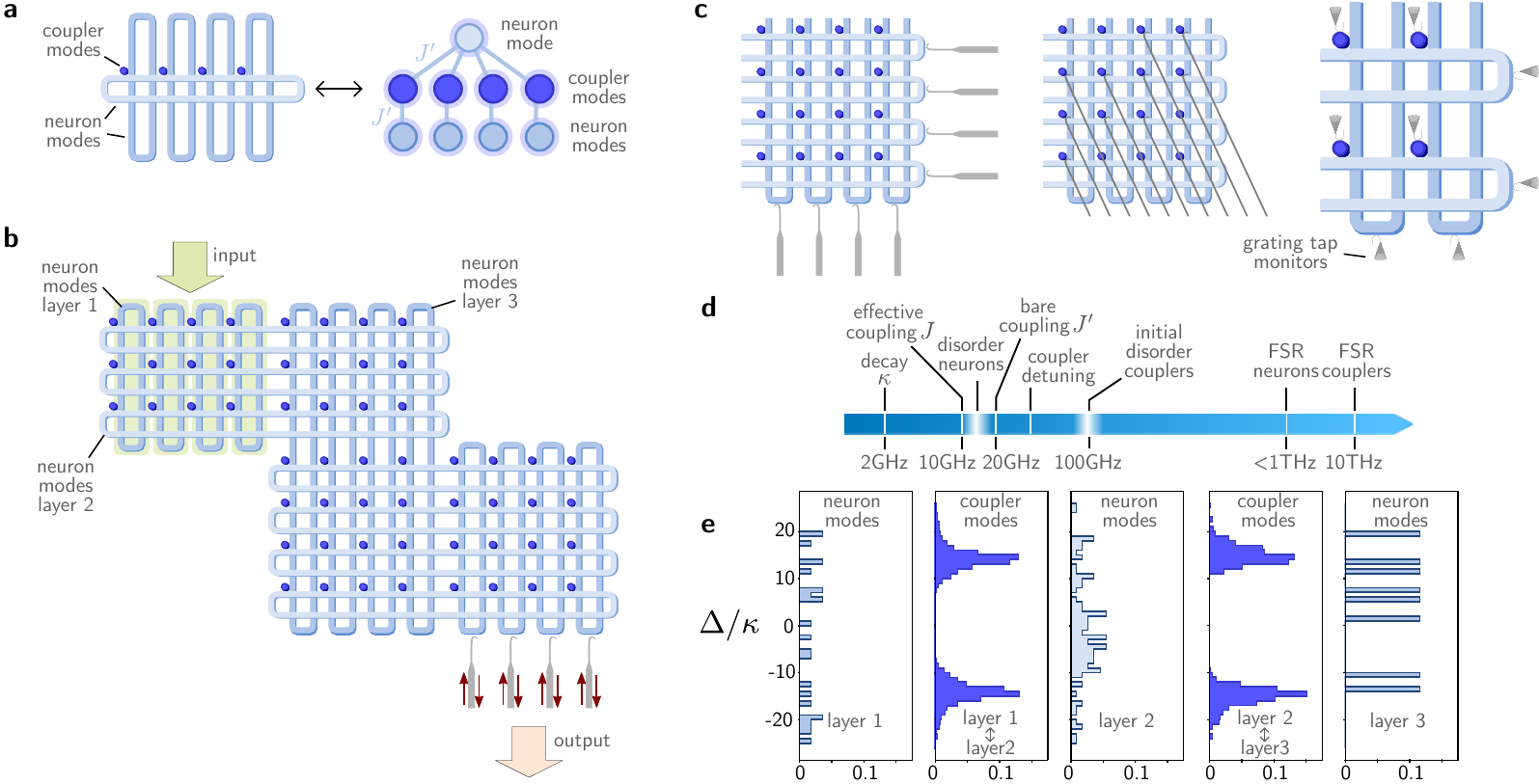}
		\caption{\textbf{Implementation with racetrack resonators.}
			\textbf{a}~Neuron modes are represented by racetrack resonators and racetrack resonators (light blue) of different layers in the neural network are crossed employing techniques to reduce the cross-talk between them. They are then coupled via microring resonators (dark blue)---the coupler modes. Changing the detuning of the coupler modes changes the effective coupling strength $J$ between any two racetrack resonators that cross. This is illustrated in the ersatz image on the right.
			\textbf{b}~The advantage of this design is that the system can be scaled up while requiring only a minimal number of waveguide crossing and the neuron modes can still be accessed with waveguides from outside.
			\textbf{c}~Three possibilities to measure gradients: following the expression for the gradient in terms of scattering matrix elements, one can either measure the response at the racetrack resonators and use Eq.~\eqref{eq:derivativeScatMatCoupling} or directly at the microring resonators and use Eq.~\eqref{eq:derivativeScatMatDetuning} to update parameters.
			Alternative to coupling to waveguides to each resonator, optical grating tap monitors can be utilised which light up according to the output signal at that resonator which can be recorded by a camera~\cite{Scarcella2016PLAT4M,pai2023experimentally}. In commercial implementations, grating tap monitors can be complemented by integrated photodetectors for a faster readout. To be sensitive to a specific quadrature~\eqref{eq:networkOutputDef}, the light coupled to the grating tap monitor can be combined with light from a local oscillator (not shown) to perform a homodyne measurement.
			\textbf{d}~Scale of the relevant frequencies in an optical implementation.
			\textbf{e}~Distribution of neuron modes and coupler mode detunings after training.
		}
		\label{fig:implementation}
	\end{figure*}%
	
	\subsection{Hyperparameters}
	The architecture we propose has the following hyperparameters which influence the accuracy and the convergence speed: (i)~the number of neuron modes per layer $N_\ell$; (ii)~the number of layers $L$; (iii)~the number of input replications $R$; (iv)~The intrinsic decay rate of the system.
	
	(i)~The number of neuron modes in the first layer should match the product of the input dimension $D$ and the number of repetitions of the input $R$. The number of neurons in the final layer is given by the output dimension, e.g., for classification tasks the dimension corresponds to the number of classes.
	The sizes of intermediate layers should be chosen to provide enough training parameters; as we can see from Fig.~\ref{fig:MNISTTraining}~\textbf{b}, both the convergence speed and the best accuracy rely on having a sufficient number of training parameters available, as for instance, a system with a second layer of only $20$ neuron modes performs worse than a system with $64$ neuron modes in the second layer.
	In particular, as we show in the SI, that independent of the architecture, the maximal number of independent couplings for an input of dimension $D$, input replication $R$ and output dimension $N_\mathrm{out}$ is given by
	\begin{align}
		N_\mathrm{opt} & = \frac{(R D + N_\mathrm{out})(R D + N_\mathrm{out}+1)}{2}.
	\end{align}
	In addition, there are $RD+N_\mathrm{out}$ local detunings that can be varied independently.
	For the layered structure of Fig.~\ref{fig:MNISTTraining}~\textbf{a}, this translates to an optimal size of the second layer $N_{2,\mathrm{opt}}=\left\lceil(N_1+N_3+1)/2\right\rceil$.
	For a system with $N_1=128$ and $N_3=10$ the optimal value $N_{2,\mathrm{opt}}$ is given by of $N_{2,\mathrm{opt}}=70$. Smaller $N_2$ leads to fewer independent parameters, while larger $N_2$ introduces redundant parameters. In our simulations, the system with $N_2=64$ already achieved the same classification accuracy as a standard artificial neural network, so it may not always be necessary to increase the layer size to $N_{2,\mathrm{opt}}$.
	Further increasing the number of parameters and introducing redundant parameters can
	help avoid getting stuck during training.
	Similarly, even though a neuromophic system with all-to-all couplings provides a sufficient number of independent parameters, considering a system with multiple hidden layers can be advantageous for the training.

	(ii)~Similar considerations apply to choosing the number of layers of the network. $L$ should be large enough to provide a sufficient number of training parameters. At the same time, $L$ should not be too large to minimise attenuation losses. For deep networks with $L\gtrapprox3$, localisation effects may become important~\cite{Iida1990Wave,Iida1990Statistical} which may hinder the training, so it is advisable to choose an architecture with sufficiently small $L$.
	
	(iii)~The choice of $R$ depends on the complexity of the training set. For the digit recognition task, $R=2$ was sufficient, but more complicated datasets can require larger $R$. We explore this question in the SI, where we show for scalar functions that $R$ determines the approximation order and utilise our system to fit scalar functions. To fit quickly oscillating functions or functions with other sharp features, we require larger $R$.
	
	Here, we only considered injecting the input at the first layer. However, it could be interesting to explore in the future whether spreading the (replicated) input over different layers holds an advantage, since this would allow to make subsequent layers more `non-linear'.
	
	(iv)~The intrinsic decay rate determines the sharpest feature that can be resolved, or, equivalently, a larger rate $\kappa$ smoothens the output functions. It is straightforward to see from Eq.~\eqref{eq:derivativeScatMatDetuning} that the derivative of the output w.r.t.~a component of the input scales with $\kappa$. 
	In the context of one-dimensional function fitting this is straightforward to picture and we provide some examples in the SI.
	However, the network does not lose its use since the larger decay rate can be compensated for by rescaling the range of the input according to $\kappa$. For instance, for the digit recognition training, we set the image background to $\Delta/\kappa=5$ and made the pixels storing the number resonant $\Delta/\kappa=0$ which is still possible in lossy systems.
	
	\section{Proposed optical implementation}
	
	\subsection{Racetrack resonator architecture}
	
	The experimental realisation of our proposed system has two main requirements: (i)~a sufficiently large number of system parameters (not necessarily all) needs to be tunable; (ii)~a large number of modes needs to be sufficiently densely connected. 
	A simple geometry could consist of localised resonators (neuron modes) connected by waveguides (couplings). However, we find a more promising geometry in terms of tunability and spatial layout.
	
	Here we propose an integrated-photonics design based on racetrack resonators as neuron modes that are coupled via microring resonators which realise an effective coupling between the neuron modes, see Fig.~\ref{fig:implementation}~\textbf{a}. This alleviates the need to tune the coupling between the neuron modes directly, since the effective coupling strength can be controlled via the detuning of the coupling resonator. The detunings of the racetrack resonators and the microring resonators could in practice, for instance, be tuned electro-optically~\cite{Guarino2007Electro,Hu2021On} or by locally heating the resonator. Alternatively, the couplings can directly be controlled electro-optically~\cite{Jia2023Electrically}.
	
	To achieve high connectivity, we propose to cross racetrack resonators of different network layers as in Fig.~\ref{fig:implementation}~\textbf{b} and use techniques developed to reduce cross-talk between overlapping waveguides~\cite{Jones2013Ultra,Johnson2020Low} to isolate the racetrack resonators of different layers from one another.
	The coupling between these neuron modes is controlled via the detuning of the microring resonators placed at the intersection between racetrack resonators which we dub \emph{coupler modes}.
	The effective coupling $J$ induced by a coupler mode of detuning $\Delta_\mathrm{coupler}$ coupled to two modes with strength $J'$ is given by $J = \lvert J'\rvert^2/(\mathrm{i}\kappa_\mathrm{coupler}/2 - (\Delta_\mathrm{coupler}-\omega))$, see the SI.
	The coupler modes also induce a frequency shift on the neuron modes by the same amount which can be compensated by alternating positive and negative detuning in neighbouring coupler modes.
	A special feature of our proposed design is that the neuron modes are spatially extended whereas the coupling is facilitated by local microring resonators.
	In fact, the device layout of Fig.~\ref{fig:implementation}~\textbf{c} can be seen as a direct visualisation of the coupling matrix $J^{(n)}$ between consecutive layers.
	In practice, it is advantageous to operate in the regime $\lvert\Delta_\mathrm{coupler}\rvert\gg\kappa_\mathrm{coupler}$ in which the effective coupling is predominantly coherent. We outline further requirements on $\Delta_\mathrm{coupler}$ and $\kappa_\mathrm{coupler}$ as well as a procedure to create a suitable initial configuration of the system in the SI.
	
	Following our procedure to measure gradients based on the scattering matrix, Eqs.~\eqref{eq:derivativeScatMatCoupling} and \eqref{eq:derivativeScatMatDetuning},
	there are three possibilities to obtain the gradients, Fig.~\ref{fig:implementation}~\textbf{c}: (i)~One only measures the scattering matrix at the racetrack resonators which directly yields the gradients according to Eq.~\eqref{eq:derivativeScatMatDetuning} w.r.t. the neuron mode detunings and uses Eq.~\eqref{eq:derivativeScatMatCoupling} to infers the gradients for the coupler modes, see the SI. This requires knowledge of the coupling strength $J'$ between coupler modes and neuron modes. (ii)~Alternatively, one directly measures the scattering response at the coupler modes and uses Eq.~\eqref{eq:derivativeScatMatDetuning} to update all parameters. This immediately yields the gradients for the detuning of the coupler modes while requiring a larger number of (crossed) waveguides. (iii)~To reduce the number of waveguides required to measure the system, one can couple grating tap monitors to each resonator which can either be monitored with a camera~\cite{pai2023experimentally} or integrated photodetectors for a faster readout. 
	To detect a specific quadrature~\eqref{eq:networkOutputDef}, the light coupled to the grating tap monitor can be combined with light from a local oscillator to perform a homodyne measurement.
	
	We simulate the system of the previous section using the ersatz description illustrated in Fig.~\ref{fig:implementation}~\textbf{a} to represent the coupler modes.
	To make the simulation as realistic as possible, we initialised the system such that the detuning range of the initially random detunings was $\lvert\Delta\rvert/\kappa\sim5$ for the neuron modes and $\lvert\Delta_\mathrm{c}\rvert/\kappa\sim10$ for the coupler modes.
	This is in-line with realistic values. We provide a complete overview of the relevant frequencies in state-of-the-art electro-optically tunable resonators in Fig.~\ref{fig:implementation}~\textbf{d}.
	We further discuss the requirements for the experimental implementation of our neuromorphic system in more detail in the next section.
	We train the system for $11,300$ iterations and achieve a 
	test accuracy of $92.6\,\%$ which is similar to the previous simplified setup. The slightly lower accuracy is due to the shorter training time.
	Fig.~\ref{fig:implementation}~\textbf{e} shows the distribution of the detunings of neuron modes and coupler modes after training. The final distributions spread over a range of approximately $\lvert\Delta\rvert/\kappa\sim\lvert\Delta_\mathrm{c}\rvert/\kappa\sim20$ which is well within the tunable range that can be realistically achieved in experiments~\cite{herrmann2023arbitrary,Zhou2020Electro}, see Fig.~\ref{fig:implementation}~\textbf{d}. We discuss requirements on the physical implementation in further detail in the next section.
	
	We propose that this layout can also find application in the implementation of other neuromorphic systems since it presents a convenient way to achieve high connectivity while keeping the number of waveguide crossings minimal.
	
	\subsection{Experimental requirements}
	
	Fig.~\ref{fig:implementation}~\textbf{d} shows the relevant frequencies that are important for the implementation of our neuromorphic system.
	The approach proposed here relies on efficient tuneability of a linear system, both for injecting the input $\textbf{x}$ and for learning. During training (or device calibration), the detuning range needs to be sufficiently large to overcome any fabrication disorder. In optical systems, disorder in the resonance frequencies typically amounts to about one percent of the free spectral range (FSR), and this can easily be overcome via electrical thermo-optic tuners that have shown \cite{armani2004electrical,gan2007maximizing} tuning by a full FSR in resonator structures of the scale considered here ($10\, \mu \mathrm{m}$ to $100 \, \mu \mathrm{m}$), see Fig.~\ref{fig:implementation}~\textbf{e}, operating on time scales \cite{gan2007maximizing} of $10 \mu \mathrm{sec}$.
	For input injection, it is desirable to have faster response times, but conversely the tuning range can be much smaller, on the order of the couplings or decay rates. This could be achieved via electro-optic tuning that has demonstrated resonator tuning by many linewidths at speeds of tens of MHz for microtoroids~\cite{baker2016high}, tens of GHz for photonic-crystal resonators~\cite{li2020lithium} and tens of GHz~\cite{herrmann2023arbitrary} for racetrack resonators.
	Tuning ranges even up to hundreds of GHz~\cite{Zhou2020Electro} were demonstrated for racetrack resonators which would be sufficient to compensate for initial frequency disorder.
	In an alternative approach to coupling the neuron modes via coupler modes, one can tune the couplings between racetrack resonators directly which has been demonstrated at multiples of the linewidth~\cite{Sacher2013Integrated}.
	Without further optimization, one can expect tuning speeds of multiple GHz~\cite{herrmann2023arbitrary,Herrmann2022Mirror} which with further optimization can be brought down to 100 GHz modulation frequency~\cite{Wang2018Integrated} allowing for fast training and inference.
	Assuming $5\,\mathrm{GHz}$ modulation frequency, one could complete all tuning steps of the entire training with $28,000$ iterations over minibatches of $200$ images reported above in only $1.1\,\mathrm{ms}$. 
	
	We would like to stress that the ideas developed here are very general and apply to a range of linear systems. In particular, we discuss in the SI how to transfer these ideas to analog electronic circuits.
	
	\section{Discussion}
	
	We introduced a concept for neuromorphic computing that relies purely on linear wave scattering and does not require any physical non-linearities. By injecting the input in the system parameters and considering the scattering response to a probe signal as output of the neuromorphic system, Fig.~\ref{fig:setup}, we obtain a non-linear function between input data and output. Replicating the input multiple times allows to realise more complex non-linear functions.
	
	With our approach, training is most straight-forward since gradients
	can be physically measured as products of scattering matrix elements which can be efficiently obtained with a minimal number of measurements.
	Furthermore, our approach does not require full knowledge of the system. In fact, we can freely add modes and couplings to the systems, which are not trained, that can be treated like a black box and the training procedure remains the same.
	
	We propose a layer architecture and gain analytic insight into its scattering matrix which has a recursive structure reminiscent of standard neural networks.
	Simulations of our neuromorphic system achieved a classification accuracy on par with the accuracy of standard artificial neural networks.
	
	Furthermore, we propose its implementation in a system of crossed, tunable racetrack resonators which are coupled through tunable microring resonators that allow to control the coupling strength. 
	We suggest that this architecture of crossed racetrack resonators can also find application in the implementation of other neuromorphic systems since it presents an efficient way to achieve a high connectivity between modes at a minimal number of waveguide crossings.
	
	Our approach solves two major challenges in the field of neuromorphic computing. We not only overcome the challenge of implementing physical non-linearities, our approach also allows to perform physical back-propagation with measured gradients.
	Moreover, our proposal can be implemented in state-of-the-art integrated and scalable, photonic circuits in which high resonator tuning speeds promise fast and reliable data processing.
	
	In future work, we propose to explore different architectures and the possibility of employing a Floquet scheme for input replication or the use of multiple modes of the same cavity to reduce the hardware overhead.
	Furthermore, the framework developed here is very general and can be applied to a range of settings beyond optics, for instance, analog electronic circuits.
	Therefore, our work opens up new possibilities for neuromorphic devices with physical back-propagation over a broad range of platforms.
	
	During the completion of our manuscript, two related works appeared as preprints~\cite{yildirim2023nonlinear,xia2023deep}.
	In contrast to these works which rely on free-space light propagation, our approach is based on the (steady state) scattering response which has potential advantages in terms of the realisable non-linear functions and allowed us to formulate a simple technique for physically extracting gradients via scattering experiments.
	
	\subsection*{Acknowledgements}
	We would like to thank Jason Herrmann for sharing useful insight into the tuning speeds of state-of-the-art electro-optical resonators. CCW would like to thank Henning Schomerus for helpful discussion.
	
	\input{main_V4_onlyMain.bbl}

	\clearpage
	
	\setcounter{figure}{0}
	\renewcommand{\thefigure}{S\arabic{figure}}
	\renewcommand{\theequation}{S\arabic{equation}}
	\setcounter{equation}{0}
	\setcounter{section}{0}
	\setcounter{table}{0}
	
	\onecolumngrid
	\part*{\large\centering Fully Non-Linear Neuromorphic Computing with Linear Wave Scattering \\ Supplementary Information}
	\begin{center}
		Clara C. Wanjura, Florian Marquardt
	\end{center}
	\twocolumngrid
	
	\section{The scattering matrix can approximate arbitrary non-linear functions}
	
	\begin{figure}
		\centering
		\includegraphics[width=.49\textwidth]{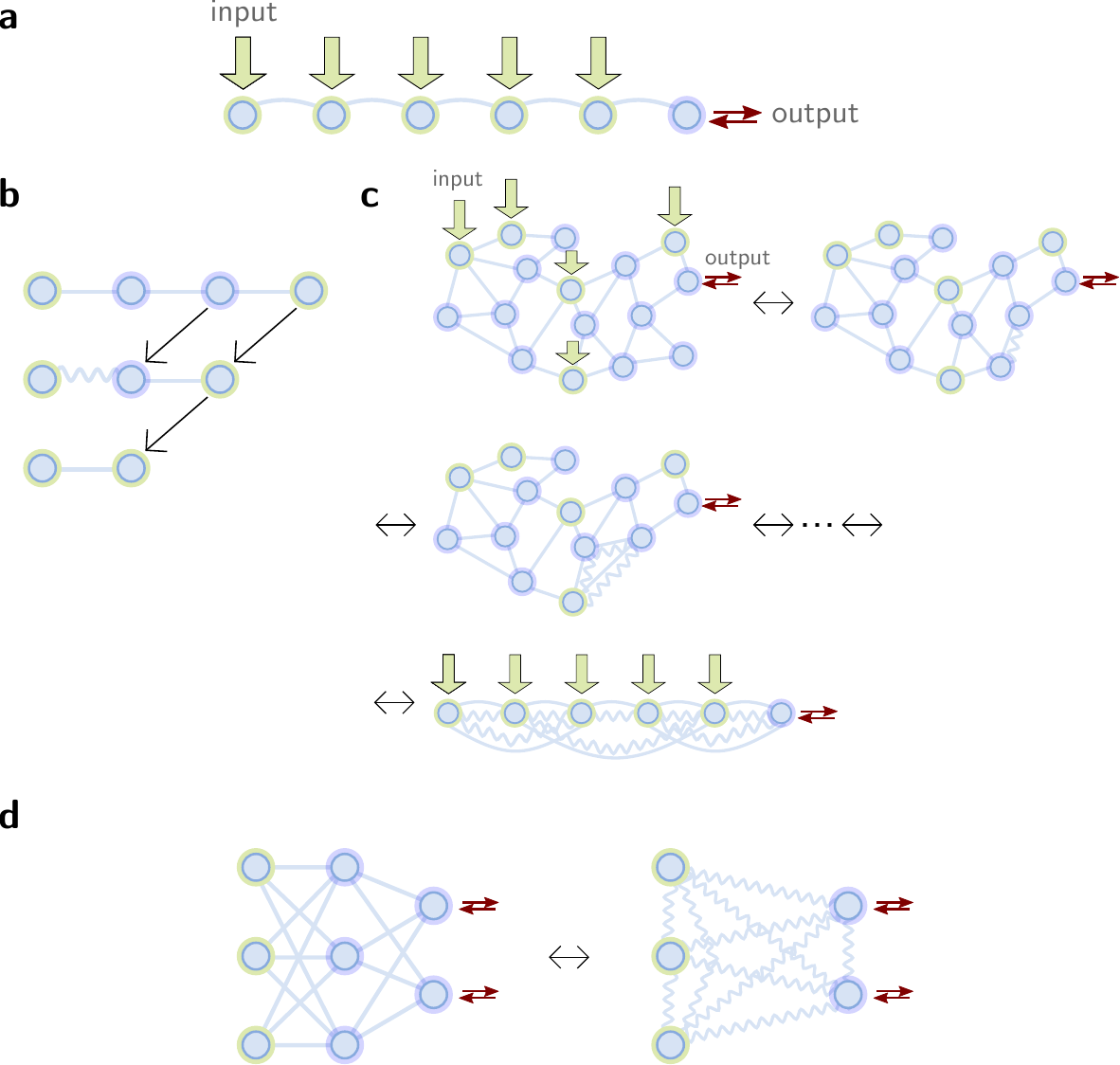}
		\caption{\textbf{Mapping of a multimode system to a one-dimensional chain.}
			\textbf{a}~The scattering matrix of a one-dimensional chain with nearest-neighbour couplings in which the input enters in the detuning of each mode has the form of a generalised continued fraction which can be used to approximate any analytic function. In the limit of an infinite chain length, this approximation approaches the exact result.
			\textbf{b}~Reduction rules to map between systems of coupled modes~\cite{Wanjura2023Quadrature}. Eliminating an intermediate mode in a chain of modes coupled via beamsplitter couplings results in effective dissipative coupling with added local decay, Eq.~\eqref{eq:singleReduction}, a further elimination step yields effective coherent beamsplitter coupling with added decay, Eq.~\eqref{eq:doubleReduction}. This technique is discussed in more detail and generalised to combinations of beamsplitter and two-mode-squeezing couplings in Ref.~\cite{Wanjura2023Quadrature}.
			\textbf{c}~Employing the reduction rules of \textbf{b}, one can map \emph{any} system of arbitrarily coupled modes to a one-dimensional chain with couplings of varying range. For clarity, not all resulting effective couplings are shown.
			\textbf{d}~Any neuromorphic system with input dimension $D$ and input replication $R$, e.g., the three-layer system of Fig.~\ref{fig:MNISTTraining}~\textbf{a}, can be mapped to a system consisting only of the input and output modes with all-to-all dissipative couplings. The number of these dissipative couplings equals the number of couplings we can at most control independently in a neuromorphic system with this input dimension and input replication.
		}
		\label{fig:mapping1dChain}
	\end{figure}%
	
	Here, we prove for the case of scalar input and output that the scattering matrix of a system of coupled modes can be used to approximate arbitrary analytic functions $x\to f(x)\in\mathbb{R}$.
	The scalar input $x\in\mathbb{R}$ is encoded in the detunings of each mode and the function is approximated by optimizing (some of) the other system parameters. In the limit of an infinitely large system, the approximation becomes exact.
	
	\subsubsection{One-dimensional chain with nearest-neighbour couplings}
	The Green's function of a one-dimensional chain with nearest-neighbour couplings is particularly straightforward to compute.
	We consider the setup for a chain of length $N=R+1$, with $R$ the number of input replications, shown in Fig.~\ref{fig:mapping1dChain}~\textbf{a}. The detuning at the $j$th mode is given by $\Delta_j+x$ for $j\leq N = R+1$ and nearest-neighbour modes $j$ and $j+1$ are coupled at strength $J_j$. The system is probed at the $N$th site where the detuning is simply $\Delta_N$.
	Similar to the derivation of the scattering matrix~\eqref{eq:ScatMatLayers} for the layered architecture, we consider the equation of motions in the frequency domain
	\begin{align}
		\omega a_j(\omega) = & \left(-\mathrm{i}\frac{\kappa_j}{2} + (\Delta_j+x)\right) a_j + J_{j-1} a_{j-1} + J_{j+1} a_{j+1}, \notag \\
		\omega a_N(\omega) = & \left(-\mathrm{i}\frac{\kappa_N+\kappa'_N}{2} + (\Delta_N+x)\right) a_N + J_{N-1} a_{N-1} \notag \\
		& - \sqrt{\kappa'_N} a_{N,\mathrm{probe}}
	\end{align}
	and recursively solve for $a_j(\omega)$ starting from $a_1(\omega)$ up to $a_N(\omega)$. Together with the input-output boundary conditions $a_{N,\mathrm{res}} = \sqrt{\kappa}a_N + a_{N,\mathrm{probe}}$, we can compute the scattering matrix element $S_{N,N}$
	\begin{widetext}
		\begin{align}\label{eq:genContinuedFraction}
			S_{N,N}(\omega,x,\theta) & = 
			1 + \frac{\kappa_N}{\mathrm{i}\frac{\kappa_{N}}{2}-(\Delta_{N}-\omega) + \frac{\lvert J_{N-1}\rvert^2}{\mathrm{i}\frac{\kappa_{N-1}}{2}-(\Delta_{N-1}+x-\omega) + \frac{\vdots}{\mathrm{i}\frac{\kappa_{2}}{2}-(\Delta_{2}+x-\omega)
						+\frac{\lvert J_{1}\rvert^2}{\mathrm{i}\frac{\kappa_{1}}{2}-(\Delta_{1}+x-\omega)}}}}.
		\end{align}
	\end{widetext}
	This scattering matrix has the form of a generalised continued fraction. More specifically, it is a so called C-type continuous fraction with the tunable parameters $J_j$ and $\Delta_j$.
	For this type of continued fraction, it has been shown that any analytic function can be expanded into a corresponding type continued fraction, short type continued fraction,~\cite{Frank1946Corresponding} which in the limit $N\to\infty$ becomes exact. Therefore, assuming that $J_j$ and $\Delta_j$ can be arbitrarily tunable, $S_{N,N}(\omega,x,\theta)$ can approximate arbitrary analytic functions with $N$ dictating the accuracy of the expansion.
	
	This expansion has a number of advantages over expansions such as power series, since it converges much faster than, for instance, a Taylor series, and it converges for any physical, dynamically stable system such as the ones we are concerned with in this work. In particular, the generalised continued fractions~\eqref{eq:genContinuedFraction} of stable physical systems are bounded at each order.
	In contrast to power series the series behaves well for large $\lvert x\rvert$ at finite $N$ since at each order, the approximated function $y(x)\to 0$ for $\lvert x\rvert\to\infty$.
	
	\subsubsection{General system of coupled modes}
	
	Let us now consider a general systems of coupled modes such as the one shown in the first panel of Fig.~\ref{fig:mapping1dChain}~\textbf{c}. The scalar input enters in the detunings $\Delta_j+x$ of $R$ sites while the other sites have detunings $\Delta_\ell$ which serve as parameters. The system is again probed at a specific site $p$ and we derive the scattering matrix element $S_{p,p}$ of that site.
	
	Here, we can employ a technique to map this system of coupled modes to a linear chain by employing a reduction technique that eliminates individual modes to map a smaller physical system with different couplings and added decay~\cite{Wanjura2023Quadrature}.
	Specifically, the rules of this reduction technique are shown in Fig.~\ref{fig:mapping1dChain}~\textbf{b}. In the first row, four modes are coupled via beamsplitter couplings
	\begin{align*}
		-\mathrm{i}\omega a_1 & = \left(-\frac{\kappa_1}{2}-\mathrm{i}\Delta_1\right) a_1 - \mathrm{i} J_1 a_2 \\
		-\mathrm{i}\omega a_2 & = \left(-\frac{\kappa_2}{2}-\mathrm{i}\Delta_2\right) a_2 - \mathrm{i} J_1^* a_1 - \mathrm{i} J_2 a_3 \\
		-\mathrm{i}\omega a_3 & = \left(-\frac{\kappa_3}{2}-\mathrm{i}\Delta_3\right) a_3 - \mathrm{i} J_2^* a_2 - \mathrm{i} J_3 a_4 \\
		-\mathrm{i}\omega a_4 & = \left(-\frac{\kappa_4}{2}-\mathrm{i}\Delta_4\right) a_4 - \mathrm{i} J_3^* a_3.
	\end{align*}
	Eliminating the third mode induces effective dissipative coupling between the second and third mode with additional local decay and a detuning shift
	\begin{widetext}
		\begin{align}\label{eq:singleReduction}
			-\mathrm{i}\omega a_2 & = \left(-\frac{\kappa_2}{2} -\mathrm{i}\Delta_2 - \frac{\lvert J_2\rvert^2}{\frac{\kappa_3}{2}+\mathrm{i}(\Delta_3-\omega)} \right) a_2 - \mathrm{i} J_1^* a_1 - \frac{J_2 J_3}{\frac{\kappa_3}{2}+\mathrm{i}(\Delta_3+\omega)} a_4 \notag \\
			-\mathrm{i}\omega a_4 & = \left(-\frac{\kappa_4}{2}-\mathrm{i}\Delta_4 \frac{\lvert J_3\rvert^2}{\frac{\kappa_3}{2}+\mathrm{i}(\Delta_3-\omega)} \right) a_4 - \frac{J_3^* J_2^*}{\frac{\kappa_3}{2}+\mathrm{i}(\Delta_3+\omega)} a_2.
		\end{align}
	\end{widetext}
	Subsequently eliminating the second mode $a_2$ yields again beamsplitter coupling with additional local decay and a detuning shift~\cite{Wanjura2023Quadrature}
	\begin{widetext}
		\begin{align}\label{eq:doubleReduction}
			-\mathrm{i}\omega a_1 & = \left(-\frac{\kappa_1}{2}-\mathrm{i}\Delta_1 - \frac{\lvert J_1\rvert^2}{-\frac{\kappa_2}{2} -\mathrm{i}(\Delta_2 - \omega) - \frac{\lvert J_2\rvert^2}{\frac{\kappa_3}{2}+\mathrm{i}(\Delta_3-\omega)}} \right) a_1 
			+\mathrm{i} \frac{J_1 J_2 J_3}{\frac{\kappa_3}{2}+\mathrm{i}(\Delta_3+\omega)} a_4
			\notag \\
			-\mathrm{i}\omega a_4 & = \left(-\frac{\kappa_4}{2}-\mathrm{i}\Delta_4 - \frac{\lvert J_3\rvert^2}{\frac{\kappa_3}{2}+\mathrm{i}(\Delta_3+\omega)} 
			+\frac{\lvert J_2 J_3\rvert^2}{\left(\frac{\kappa_3}{2}+\mathrm{i}(\Delta_3+\omega)\right)^2}
			\right) a_4 +\mathrm{i} \frac{J_3^* J_2^*J_1^*}{\frac{\kappa_3}{2}+\mathrm{i}(\Delta_3-\omega)} a_1.
		\end{align}
	\end{widetext}
	The reduction approach can be applied in large networks of modes in which the elimination step of a mode involves replacing all couplings this mode was connected to by the appropriate coherent or dissipative coupling as described above.
	This technique is discussed in more detail and generality in the Supplementary Information of Ref.~\cite{Wanjura2023Quadrature} where also the generality of the argument to two-mode and single-mode squeezing is derived; a related technique is employed in Ref.~\cite{Ranzani2015Graph,Naaman2022Synthesis}.
	
	For the following arguments, the values of the effective coupling strength, effective local decay and detuning shifts are not relevant. We merely employ this reduction technique to map one physical system to one that is easier to analyse.
	In particular, we map the generic system of multiple modes, such as the first panel in Fig.~\ref{fig:mapping1dChain}~\textbf{c}, 
	to a chain with arbitrary, possibly long-range couplings, such as the chain in the last panel of Fig.~\ref{fig:mapping1dChain}~\textbf{c}, consisting only of the modes whose detunings encode the input and the mode $p$ which is used to probe the system which is the $(L+1)$th modes of the chain. 
	This is possible by performing multiple reduction steps eliminating all other modes as is sketching in the other panels of Fig.~\ref{fig:mapping1dChain}~\textbf{c}. The couplings between the modes of this reduced model are now a combination of coherent beamsplitter coupling and dissipative coupling, with coupling constants which may even be complex. We introduce the effective coupling strengths $g_{j,\ell}$ for the effective coupling between modes $j$ and $\ell$ in which we indexed modes from left to right. We note that $g_{j,\ell}$ need not be symmetric, i.e., $g_{j,\ell}\neq g_{\ell,j}^*$. The on-site terms, i.e., decay rate and detuning, were also modified by the reduction to the chain and we define $\mu_j\equiv-\frac{\tilde\kappa_j}{2}-\mathrm{i}\Delta_j$ as the effective on-site term of the $j$th site in the equations of motion for the reduced chain.
	We can then proceed as before for the chain with nearest-neighbour couplings to derive the scattering matrix element at the $(R+1)$th site, $S_{L+1,L+1}=S_{p,p}$ by eliminating mode by mode along the chain
	\begin{align}\label{eq:SMatContinuedFractionGeneral}
		S_{R+1,R+1}(\omega, x, \theta) & = 1 + \frac{\kappa_{R+1}}{\tilde\mu_{R+1}}
	\end{align}
	with
	\begin{align}
		\tilde\mu_j & = \mu_j + \frac{g_{j,j-1}g_{j-1,j}}{\tilde\mu_{j-1}} + \sum_{\substack{\text{paths of}\\\text{length }\ell:\\ 3\leq\ell\leq 2^j}}
		(-\mathrm{i})^\ell \frac{g_{j,m_1}g_{m_1,m_2}\dots g_{m_\ell,j}}{\tilde\mu_{m_1}\tilde\mu_{m_2}\dots\tilde\mu_{m_\ell}}
	\end{align}
	with $\tilde\mu_1=\mu_1$.
	Here, the sum goes over all paths from mode $j$ to modes $m_1, m_2,\dots, m_\ell<j$ and come back to mode $j$ that have a length smaller than $2^j$ and involve at least one coupling of a range longer than one, i.e., next-nearest neighbour coupling and beyond. Note that the scalar input $x$ enters as summand in each $\mu_j$.
	
	As result of this exercise, we see that Eq.~\eqref{eq:SMatContinuedFractionGeneral} reduces to Eq.~\eqref{eq:genContinuedFraction} in the case of only symmetric nearest-neighbour couplings in which case we know that $S_{R+1,R+1}(\omega,x,\theta)$ can represent arbitrary, non-linear analytic functions. In the general case, the expression for $S_{R+1,R+1}(\omega,x,\theta)$, Eq.~\eqref{eq:SMatContinuedFractionGeneral}, contains sums of generalised continued fractions in which the coefficients are again generalised continued fractions. So Eq.~\eqref{eq:SMatContinuedFractionGeneral} has both more parameters and contains more general terms, e.g., powers of $x$, so is by no means more restrictive than expression~\eqref{eq:genContinuedFraction} which already allowed to approximate arbitrary analytic functions.
	We therefore conclude that a scattering matrix of any linear system in which the input enters $R$ times can be used to approximate arbitrary analytic functions. The number of input replications $L$ determines the recursive depth of expression~\eqref{eq:SMatContinuedFractionGeneral} and therefore plays a crucial role for the accuracy of the approximation.
	
\begin{figure*}
		\centering
		\includegraphics[width=.98\textwidth]{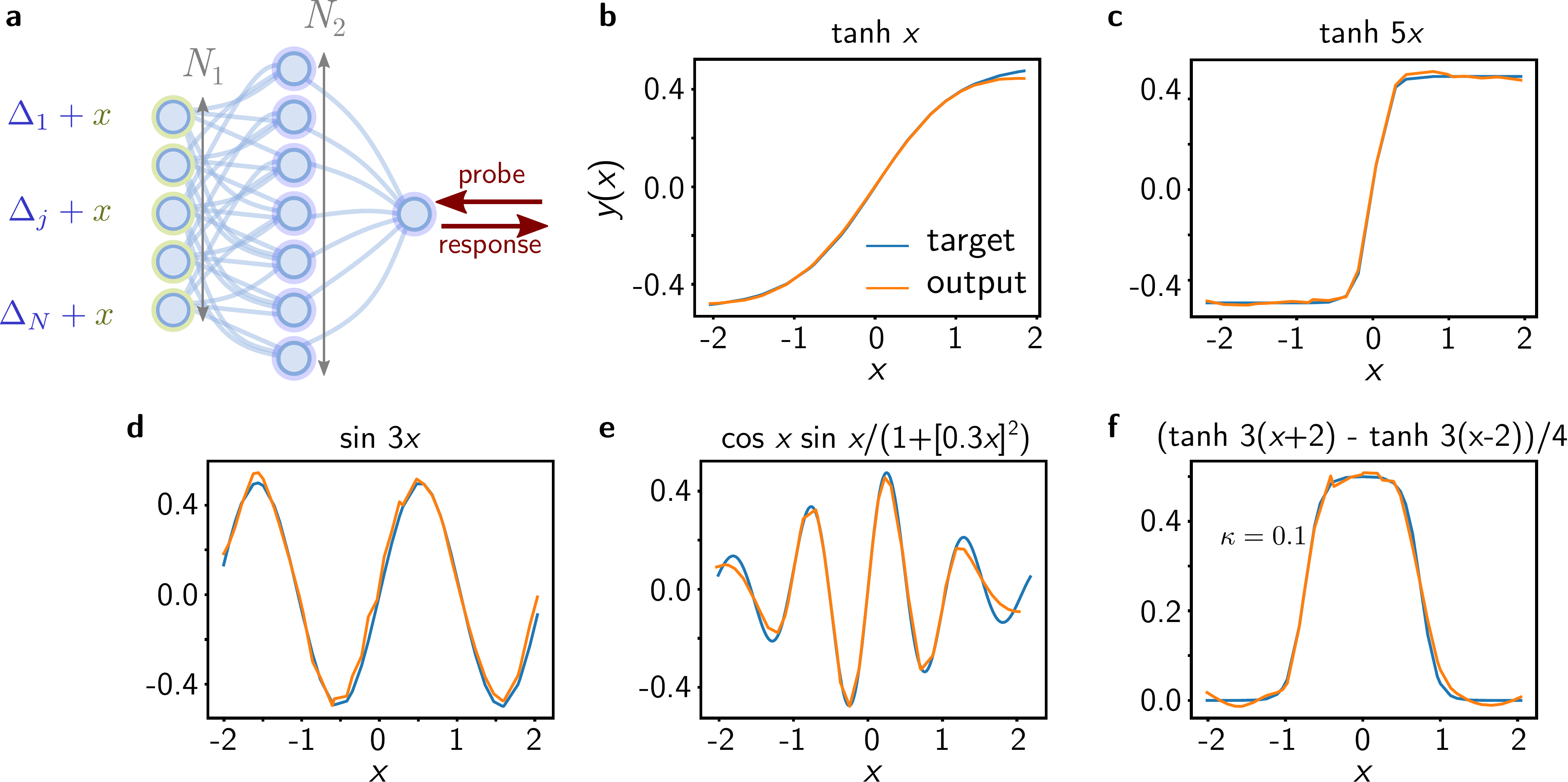}
		\caption{\textbf{One-dimensional function fitting.}
			Examples for one-dimensional functions fitted with the network shown in \textbf{a}. The input enters as offset to the detuning of the first layer and enters $R=N_1$ number of times. A second layer of variable size is connected to a single site. The imaginary part of the scattering response at that site serves as output.
			The parameters used in the fits were
			\textbf{b}~$N_1=20, N_2=20, \kappa=1$; \textbf{c}~$N_1=60, N_2=30, \kappa=0.2$; \textbf{d}-\textbf{f}~$N_1=60, N_2=30, \kappa=0.1$.
			Note that functions with sharper features require larger $R$ and larger $N_2$. Typically, $\kappa$ has to be small to capture sharp features, although, this can to some extent be compensated by rescaling the range of input values by $1/\kappa$.
		}
		\label{fig:Training1D}
	\end{figure*}

	To exemplify the expressive power of the scattering matrix, we employ the system sketched in Fig.~\ref{fig:Training1D}~\textbf{a} to fit different non-linear functions ranging from a combination of hyperbolic tangents to oscillating functions.
	The system we consider consists of a first layer with $R$ modes in which the input is replicated $R$ times in the detunings $\Delta_j^{(1)}+x$ of the first layer, then a second layer of variable size and an output layer with only one mode. The decay rate $\kappa$ is considered equal at all sites.
	
	We note that functions with sharper features or many oscillations require larger $R$ than smoother functions. Furthermore, we selected for each plot the largest value of $\kappa$ that still yielded a good approximation of the function. As we would expect from Eq.~\eqref{eq:derivativeScatMatDetuning} of the main text, a larger derivative requires a smaller decay rate $\kappa$.

	\section{Expressibility}
	
	\subsubsection{Number of independent parameters}
	In general, a scattering matrix element is of the form
	$S_{j,\ell}(0) = 1+\mathrm{i}\kappa_j (-1)^{j+\ell}\frac{\det H^{(j,\ell)}}{\det H}$.
	For a neuromorphic system with multi-dimensional input and output, both numerator and denominator automatically contains terms such as $x_1 x_2$, $x_2 x_3$ and $x_1 x_2 x_3$ even at $R=1$ with some coefficients defined through the other system parameters. In fact, with $D$ the input dimension and $R$ the number of input replications, the number of these terms grows as
	\begin{align}
		\sum_{j=1}^{DR} \binom{DR}{j} = 2^{DR} - 1.
	\end{align}
	However, while, in general, the number of terms grows exponentially with the input dimension and input replication, not all of these coefficients are independent parameters that can be controlled by adjusting the other system parameters.
	The number of independent parameters is typically much smaller.
	
	To find the actual number of independent parameters we can maximally achieve for a neuromorphic system with input dimension $D$, input replication $R$ and output dimension $N_\mathrm{out}$, including systems with the three-layer architecture of Fig.~\ref{fig:MNISTTraining}~\textbf{a}, we consider the map to the minimal reduced system which still contains all input sites and probe sites, see Fig.~\ref{fig:mapping1dChain}~\textbf{d}.
	The number of couplings entering in this system is equal to the maximal number of independent parameters we can maximally achieve for this given input dimension $D$ and input replication $R$.
	We find that the minimal system consists only of input and output modes with all-to-all dissipative couplings between them, 
	see Fig.~\ref{fig:mapping1dChain}~\textbf{d}. 
	The total number of couplings in this system is
	\begin{align}\label{eq:numberCouplingsEquivalentSystem}
		N_\mathrm{max} = 
		\frac{(DR + N_\mathrm{out})(DR + N_\mathrm{out} + 1)}{2}.
	\end{align}
	In the three-layer architecture, $N_1=D R$ and $N_3 = N_\mathrm{out}$.
	In addition to that, we have $D R + N_\mathrm{out}$ local terms (decay rates, detunings).
	Expression~\eqref{eq:numberCouplingsEquivalentSystem} is the maximal number of couplings we can at most control independently in any system at given $D$, $R$ and $N_\mathrm{out}$ independent of the architecture.
	
	Next, we come back to the three-layer architecture, Fig.~\ref{fig:MNISTTraining}~\textbf{a}, and we ask how large the second layer $N_2$ has to be to have enough independent parameters to control all of these effective couplings independently. (The $N_1 + N_3$ local parameters are automatically independent parameters since they also exist in the original system.)
	The total number of couplings between layers of sizes $N_1$, $N_2$ and $N_3$ is given by
	\begin{align}\label{eq:numberCouplings}
		N_\mathrm{control} = N_1 N_2 + N_2 N_3.
	\end{align}
	This is the number of couplings, we can actually control.
	When $N_\mathrm{control}<N_\mathrm{max}$ we do not utilise the full potential of the architecture, whereas when $N_\mathrm{control}>N_\mathrm{max}$, some of the couplings are, at least in principle, redundant and the system has a hardware overhead.
	
	The optimal case is when $N_\mathrm{control}=N_\mathrm{max}$. Equating expressions~\eqref{eq:numberCouplingsEquivalentSystem} and \eqref{eq:numberCouplings}, we find that to be able to independently control all parameters, $N_2$ should be larger or equal to
	\begin{align}
		N_2 & \geq 
		\left\lceil
		\frac{N_1+N_3+1}{2}
		\right\rceil.
	\end{align}
	For the system we consider in the main text with $N_1=128$ and $N_3=10$, we find $N_2\geq 70$.
	This explains why the classification accuracy is limited for smaller $N_2$ and why we achieve the highest classification accuracy for $N_2=64$.
	
	Depending on the training data, it may also be worth to explore larger $N_2$ than this bound, even though the system is then over determined, since this may have advantages during training.
	
	\subsubsection{Remarks about systems encoding functions with multi-dimensional input and output}
	
	In the main text, we derive the expression for the scattering matrix~\eqref{eq:ScatMatLayers} of a neuromorphic system with a layered architecture, Fig.~\ref{fig:MNISTTraining}~\textbf{a}. The mathematical form of this equation resembles that of the generalised continued fraction~\eqref{eq:genContinuedFraction} but in matrix form.
	
	In the literature of condensed matter physics, expressions such as Eq.~\eqref{eq:ScatMatLayers} are known as recursive Green's functions~\cite{Sols1995Recursive} and have been employed to study various condensed matter systems. It is expected that for larger depth $L$, these networks display localisation when subject to disorder in the couplings and detunings~\cite{Iida1990Wave,Iida1990Statistical}---this may possibly impede the training of such networks (correlations between input and output layer are expected to decrease with $L$ in which case the gradients will also become small). For most applications, $L=3$ or $4$ may already be sufficient.
	If the application requires deeper networks, introducing gain, e.g., in the form of two-mode or single-mode squeezing, or, alternatively, introducing a few strategic couplings of longer range between layers that are further apart, may resolve this issue of localisation.

	\section{Derivation of the gradient formulas}
		Here, we derive the derivative of the scattering matrix w.r.t. a parameter $\theta_j$, Eq.~\eqref{eq:SMatDerivativeGeneral},
		\begin{align}\label{eq:ScatMatDerivativeDerivation}
			\frac{\partial S(\omega,\mathbf{x},\theta)}{\partial \theta_j}
			& = - \mathrm{i}\sqrt{\kappa} \frac{\partial}{\partial \theta_j} [\omega\mathbb{1}-H(\mathbf{x},\theta)]^{-1} \sqrt{\kappa}.
		\end{align}
		The derivative of the matrix inverse $M^{-1}$ w.r.t. a parameter follows from
		\begin{align}
			0 & = \frac{\partial}{\partial \theta_j}\mathbb{1}
			= \frac{\partial}{\partial \theta_j} \left(M^{-1}(\theta) M(\theta)\right) \notag\\
			& = \left(\frac{\partial}{\partial \theta_j} M^{-1}(\theta)\right)M(\theta) + M^{-1}(\theta)\left(\frac{\partial}{\partial \theta_j} M(\theta)\right).
		\end{align}
		Solving for $\frac{\partial}{\partial \theta_j} M^{-1}(\theta)$, we obtain
		\begin{align}
			\frac{\partial}{\partial \theta_j} M^{-1}(\theta)
			= M^{-1}(\theta)\left(\frac{\partial}{\partial \theta_j} M(\theta)\right)M^{-1}(\theta).
		\end{align}
		Using this expression for Eq.~\eqref{eq:ScatMatDerivativeDerivation}, we obtain Eq.~\eqref{eq:SMatDerivativeGeneral} of the main text.
	
	\section{Further details about the implementation}
	
	\subsection{Derivation of the effective coupling}
	\begin{figure}
		\centering
		\includegraphics[width=.49\textwidth]{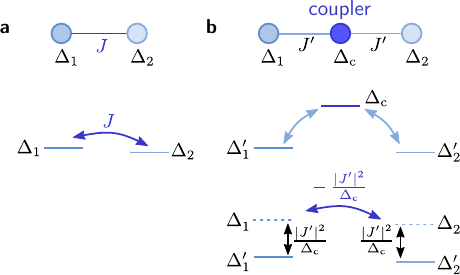}
		\caption{\textbf{Effective coupling mediated by coupler modes.}
			The aim of the implementation, Fig.~\ref{fig:implementation}~\textbf{a}, is to replace \textbf{a}~the tunable couplings of strength $J$ by \textbf{b}~a coupler mode at tunable detuning $\Delta_\mathrm{c}$ coupled to two neuron modes at fixed coupling strength $J'$. This gives rise to an effective coupling between the two neuron modes at strength $J=\lvert J'\rvert^2/\big(\mathrm{i}\frac{\kappa_\mathrm{c}}{2}+\omega-\Delta_\mathrm{c}\big)$, Eq.~\eqref{eq:eomsNeuronCouplerModesReduced}, so the effective coupling strength $J$ can be tuned by tuning $\Delta_\mathrm{c}$. At the same time, the detuning of each mode is shifted by the same amount. To map back to exactly the same situation as in \textbf{a}, this shift needs to be compensated when tuning $\Delta_j'$, Eq.~\eqref{eq:detuningShift}.
		}
		\label{fig:effectiveCoupling}
	\end{figure}
	In Fig.~\ref{fig:implementation} we introduced coupler modes mediating the coupling between two neuron modes. Here, we derive how a change in the detuning of the coupler mode changes the effective coupling between two neuron modes.
	The aim is to achieve an effective, predominantly coherent coupling between two neuron modes which can be tuned by adjusting the detuning of the coupler mode, see Fig.~\ref{fig:effectiveCoupling}. This allows to transfer all results obtained for a network with tunable coherent coupling strengths to this system with coupler modes.
	
	For simplicity consider the two neuron modes $a_1$ and $a_2$ at detunings $\Delta_1$ and $\Delta_2$, respectively, coupled via coupler mode $c$ at detuning $\Delta_\mathrm{c}$ and experiencing decay at $\kappa_\mathrm{c}$ to which both $a_1$ and $a_2$ are coupled at strengths $J'2$. The corresponding, Fourier transformed equations of motion are given by
	\begin{align}\label{eq:eomsNeuronCouplerModes}
		-\mathrm{i}\omega a_1 & = -\left(\frac{\kappa}{2} + \mathrm{i}\Delta_1'\right) a_1 - \mathrm{i} J' c \notag \\
		-\mathrm{i}\omega a_2 & = -\left(\frac{\kappa}{2} + \mathrm{i}\Delta_2'\right) a_2 - \mathrm{i} (J')^* c \notag \\
		-\mathrm{i}\omega c & = -\left(\frac{\kappa_\mathrm{c}}{2} + \mathrm{i}\Delta_\mathrm{c}\right) c - \mathrm{i} (J')^* a_1 - \mathrm{i} J' a_2.
	\end{align}
	Solving for $c(\omega)$, we find
	\begin{align}
		c & =  \frac{J'}{\mathrm{i}\frac{\kappa_\mathrm{c}}{2} + \omega - \Delta_\mathrm{c}} (a_1 + a_2)
	\end{align}
	which we insert in Eqs.~\eqref{eq:eomsNeuronCouplerModes}
	\begin{align}\label{eq:eomsNeuronCouplerModesReduced}
		-\mathrm{i}\omega a_1 & = -\left(\frac{\kappa}{2} + \mathrm{i}\Delta_1'\right) a_1 - \mathrm{i} \frac{\lvert J'\rvert^2}{\mathrm{i}\frac{\kappa_\mathrm{c}}{2} + \omega - \Delta_\mathrm{c}} (a_1 + a_2) \notag \\
		-\mathrm{i}\omega a_2 & = -\left(\frac{\kappa}{2} + \mathrm{i}\Delta_2'\right) a_2 - \mathrm{i} \frac{\lvert J'\rvert^2}{\mathrm{i}\frac{\kappa_\mathrm{c}}{2} + \omega - \Delta_\mathrm{c}} (a_1 + a_2).
	\end{align}
	There are two key points to notice: (i)~the coupler mode induces effective coupling between the two modes in which the coupling strength is complex, i.e., it results in both coherent and dissipative coupling.
	To come as close as possible to an effectively coherent coupling, we require $\kappa_\mathrm{c}\ll\Delta_\mathrm{c}$, in which case the coherent coupling strength 
	is approximately $-\lvert J'\rvert^2/(\Delta_\mathrm{c}-\omega)$. 
	(ii)~the coupler modes lead to an on-site shift in detuning and contributes towards the dissipation.
	To compensate for this detuning shift, we offset $\Delta_j'$ by the same amount, i.e.,
	\begin{align}\label{eq:detuningShift}
		\Delta_j' & = \Delta_j + \lvert J'\rvert^2/\Delta_\mathrm{c}.
	\end{align}
	This gives rise to
	\begin{align}
		-\mathrm{i}\omega a_1 & = -\left(\frac{\kappa}{2} + \mathrm{i}\Delta_1\right) a_1 + \mathrm{i} \frac{\lvert J'\rvert^2}{\Delta_\mathrm{c}} a_2 \notag \\
		-\mathrm{i}\omega a_2 & = -\left(\frac{\kappa}{2} + \mathrm{i}\Delta_2\right) a_2 + \mathrm{i} \frac{\lvert J'\rvert^2}{\Delta_\mathrm{c}} a_1
	\end{align}
	which are the equations describing the scenario depicted in Fig.~\ref{fig:effectiveCoupling}~\textbf{a} with $J=-\lvert J'\rvert^2/\Delta_\mathrm{c}$
	Note that if the neuron modes are coupled to multiple coupler modes, the detuning shifts from all coupler modes have to be compensated, so, more generally,
	\begin{align}\label{eq:detuningShiftSum}
		\Delta_j' & = \Delta_j + \sum_\ell \lvert J'\rvert^2/\Delta_{\ell,\mathrm{c}}
	\end{align}
	in which $\ell$ sums over all coupler modes that mode $j$ is connected to.
	
	In practise, a neuron mode can be connected to a large number of coupler modes, so to avoid large detuning shifts (which may then even be outside the tuning range), one can alternate positive and negative detuning of the coupler modes. In that case, most of the terms in Eq.~\eqref{eq:detuningShiftSum} balance out and the net shift is very small.
	
	\subsection{Creating the initial configuration in the proposed implementation}
		To start training the neuromorphic system, it is first necessary to calibrate the setup, so any fabrication disorder should not be larger than the tunable range, see Fig.~\ref{fig:implementation}~\textbf{d}.
		Once the setup is calibrated, one can fine tune the detunings to a suitable initial configuration.
		We found in our simulations that starting from configurations in which the neuron modes are effectively approximately resonant (up to some small disorder), i.e., taking into account the detuning shifts~\eqref{eq:detuningShiftSum} due to the coupler modes, leads to a faster and reliable convergence of the training.
		The detunings of the coupler modes should be large enough to enable sufficiently strong coupling between neuron modes.
		In particular, $\Delta_\mathrm{c}/\kappa_\mathrm{c}\gg 1$ ensures that the coherent coupling dominates over the dissipative coupling induced by the coupler mode, Eq.~\eqref{eq:eomsNeuronCouplerModes}.
		In addition, some small disorder in the effective coupling strengths helps to avoid getting stuck early in the training.
		
		Concretely, in numerical experiments, it has proven helpful to generate a random initial configuration in the ersatz description, i.e., without the coupler modes and to translate this back to a description in terms of coupler modes, i.e., solving for $\Delta_\mathrm{c}$. Subsequently, one chooses alternating sign for the coupler mode detunings and calculates the resulting on-site shift~\eqref{eq:detuningShiftSum} which is then compensated for by changing the neuron mode detuning which effectively makes the neuron modes resonant again.
		This serves as initial configuration for the training. Subsequent training steps can then be performed \emph{in sito} based purely on the measured scattering matrix.
	
	\section{Analog electrical circuits}
	The general idea of exploiting the non-linear structure of the Green's function of a linear system transfers to other systems such as analog electrical circuits. Here, we sketch the idea for a circuit network of resistors, but the same principle holds for more general electric circuits that can, for instance, also contain inductors and capacitors.
	
	Consider an electrical circuit in which resistors $\rho_{j,\ell}$ connect nodes $j$ and $\ell$ in a resistor network. Kirchhoff's law establishes a relation between the current $I_n$ at each node $n$ and the voltage $V_n$ via the connectivity matrix $L$ or Laplacian of the network. Concretely,
	\begin{align}
		-I_n & = \sum_m L_{n,m} V_m.
	\end{align}
	with
	\begin{align}
		L_{m,n} = \begin{cases}
			\frac{1}{\rho_{m,n}} & : m\neq n \\
			-\sum_m \frac{1}{\rho_{m,n}} & : m = n
		\end{cases}.
	\end{align}
	Our ultimate goal is to calculate the resistance between nodes $j$ and $\ell$.
	To that purpose, we solve the above equation for the voltages, by calculating the circuit's Green's function~\cite{Wu2004Theory,Cernanova2014Non,Cserti2011Uniform}
	\begin{align}
		G & = (L - \mathbf{v} \cdot \mathbf{v}^\mathrm{T})^{-1}
	\end{align}
	with $\mathbf{v}=(1,1,\cdots,1)^\mathrm{T}$. This matrix of ones $\mathbf{v} \cdot \mathbf{v}^\mathrm{T}$ is the outer product of the eigenvector corresponding to the zero eigenvalue of $L$ which accounts for the fact that adding a constant, uniform potential to every node yields the same circuit with the same relative voltages.
	
	The two-point resistance $R_{j,\ell}$ between nodes $j$ and $\ell$ is then given through elements of this Green's function
	\begin{align}
		R_{j,\ell} & = G_{j,j} + G_{\ell,\ell} - (G_{j,\ell} + G_{\ell,j}).
	\end{align}
	The elements $R_{j,\ell}$ define the \emph{impedance matrix} of the system.
	Similar as for the scattering matrix, the impedance matrix is a non-linear function of the resistors $\rho_{m,n}$, so we can interpret a set of tunable $\rho_{m,n}$ to be the input $\mathbf{x}$ of our neuromorphic system and some other tunable resistors $\rho_{m',n'}$ to be the learning parameters $\theta$. The output is the measured two-point resistance $R_{j,\ell}$ between a suitable set of nodes.
	If the ground of the circuit is fixed at a certain node, e.g.~node $j$, the output is proportional to the measured voltage difference to nodes $\ell$ such that the output can be recorded with a single measurement step.
	As before, the linear nature of the system enables us to directly measure gradients via the impedance matrix.


	\input{Supplementary.bbl}

\end{document}

%% file: main_V4_onlyMain.bbl
%

%% file: Supplementary.bbl
%